# Subspecialty-Specific Foundation Model for Intelligent Gastrointestinal Pathology

Lianghui Zhu[1,2,3,4#], Xitong Ling[1,2#], Minxi Ouyang[1,4#], Xiaoping Liu[2#], Tian Guan[1], Mingxi Fu[1], Zhiqiang Cheng[5], Fanglei Fu[1], Maomao Zeng[6], Liming Liu[3], Song Duan[7], Qiang Huang[8], Ying Xiao[9], Jianming Li[10], Shanming Lu[11], Zhenghua Piao[12], Mingxi Zhu[1], Yibo Jin[13], Shan Xu[10], Qiming He[1], Yizhi Wang[1], Junru Cheng[14], Xuanyu Wang[2], Luxi Xie[4]*, Houqiang Li[3]*,Sufang Tian[2]*, Yonghong He[1]*

1. Institute of Biopharmaceutical and Health Engineering, Tsinghua Shenzhen International Graduate School, Shenzhen, Guangdong 518055, China.

2. Department of Pathology, Zhongnan Hospital of Wuhan University, Wuhan, Hubei Province 430000, China.

3. Department of Pathology, Fuzhou University Affiliated Provincial Hospital, Fuzhou, Fujian 350001, China.

4. Department of Pathology, Liuzhou People's Hospital Affiliated to Guangxi Medical University, Liuzhou, Guangxi 545001, China.

5. Department of Pathology, the Second Affiliated Hospital of Southern University of Science and Technology, Shenzhen, Guangdong 518112, China.

6. Shenzhen Zhengjingda Instrument Co., Ltd., Shenzhen, Guangdong 518101, China.

7. Department of Pathology, Chongqing University Affiliated Three Gorges Hospital, Chongqing 404000, China.

8. Shenzhen Shengqiang Technology Co., Ltd., Shenzhen, Guangdong 518122, China.

9. Department of Pathology, Beijing Tsinghua Changgung Hospital, School of Clinical Medicine, Tsinghua University, Beijing 100041, China.

10. Provincial Key Laboratory for Precision Pathology and Intelligent Diagnosis, The First Affiliated Hospital, Jiangxi Medical College, Nanchang University, Nanchang, Jiangxi 330000, China.

11. Longgang Central Hospital of Shenzhen, Shzen, Guangdong 518116, China.

12. Department of Pathology, Ningbo Clinical Pathology Diagnosis Center, Ningbo, Zhejiang 315031, China

13. School of Foreign Studies, Guangzhou University, Guangzhou, Guangdong 510006, China.

14. Medical Optical Technology R&D Center, Research Institute of Tsinghua, Pearl River Delta, Guangzhou 510700, China.

*: Corresponding author.

#: These authors contribute equally.

## Abstract

Gastrointestinal (GI) diseases represent a clinically significant burden, necessitating precise diagnostic approaches to optimize patient outcomes. Conventional histopathological diagnosis suffers from limited reproducibility and diagnostic variability. To overcome these limitations, we develop Digepath, a specialized foundation model for GI pathology. Our framework introduces a dual-phase iterative optimization strategy combining pretraining with fine-screening, specifically designed to address the detection of sparsely distributed lesion areas in whole-slide images. Digepath is pretrained on over 353 million multi-scale images from 210,043 H&E-stained slides of GI diseases. It attains state-of-the-art performance on 33 out of 34 tasks related to GI pathology, including pathological diagnosis, protein expression status prediction, gene mutation prediction, and prognosis evaluation. We further translate the intelligent screening module for early GI cancer and achieve near-perfect 99.70% sensitivity across nine independent medical institutions. This work not only advances AI-driven precision pathology for GI diseases but also bridge critical gaps in histopathological practice.

## Introduction

The global burden of gastrointestinal (GI) malignancies, particularly gastric and colorectal cancers, continues to rise significantly[1–3]. According to World Health Organization (WHO) statistics, GI cancers rank among the most prevalent malignancies worldwide, with mortality rates remaining alarmingly high in many regions[4,5]. Early detection and treatment of GI tumors are critical for improving patient survival rates and clinical outcomes[6–9]. Advances in medical imaging and pathology have enhanced the importance of early screening and precise diagnosis for GI diseases[10–12]. Clinically, biopsies remain essential for lesion characterization, inflammation grading, and tumor classification[13,14]. Endoscopic submucosal dissection (ESD) is indicated for precancerous lesions and early-stage cancers, while surgical resection remains the standard approach for advanced or undifferentiated tumors[15–18]. Postoperative pathological evaluation provides comprehensive tumor characterization, including histological type, tumor dimensions, invasion depth, pathological staging, and molecular profile[19–23]. These critical parameters facilitate accurate assessment of treatment efficacy and reliable prediction of patient prognosis[22,23]. However, conventional histopathological assessment remains limited by interobserver variability, especially in large-scale screening for early-stage malignancies[24–26].

Artificial intelligence (AI) holds significant promise for both alleviating pathologists' workload and improving diagnostic precision[27–29]. Recently, the advent of foundation models has substantially increased the potential for AI deployment in clinical pathology practice. Through self-supervised learning on millions of whole-slide images (WSIs), researchers have trained vision transformer (ViT[30]) with hundreds of millions of parameters. These foundation models have demonstrated exceptional versatility across multiple downstream tasks, including tissue type classification, tumor segmentation, genomic mutation prediction, and prognostic analysis[31–35]. Remarkably, UNI, pretrained on 100 million hematoxylin and eosin (H&E)-stained slides across 20+ tissue types, outperformed existing models on 34 diagnostic tasks[36]. Gigapath established a hierarchical feature encoding architecture for multi-scale pathological feature representation, achieving state-of-the-art performance in 25 out of 26 benchmark cancer diagnostic tasks after pretraining on 1.3 billion WSIs[37]. Subsequently, TITAN pioneered vision-language alignment paradigm for slide-level representation learning and enhanced the feature embedding power for WSI[38].

These studies confirm that foundation models pretrained on large-scale pathological datasets significantly outperform conventional models (e.g., ImageNet-pretrained and TCGA-derived architectures) in accuracy, sensitivity, and generalizability. Despite these advances, their performance still requires further enhancement for reliable clinical adoption across various specialized diagnostic tasks. Current foundation models predominantly train on pathology images at a single magnification level, whereas real-world clinical practice necessitates multi-scale analysis[39–41]. For instance, grading of atrophy and intestinal metaplasia is typically performed at low magnification (e.g., 5×), while assessment of acute inflammatory activity requires high-power examination (e.g., 20×). Moreover, for AI-based early cancer screening to be clinically available, it must achieve near-perfect sensitivity, minimizing false negatives without compromising an acceptable false-positive rate[42]. While fine-tuning feature strategies for specific downstream tasks have been proposed to enhance model performance[43–46], such approaches often compromises the generalizability of the model. In practical GI pathology workflows, where diverse diagnostic tasks coexist, maintaining multiple task-specific feature extractors would: (1) introduce prohibitive computational overhead during task-switching, (2) necessitate costly hardware infrastructure to store and run parallel large-scale feature encoders, and (3) offer no guarantee of model generalizability.

To overcome these limitations, we developed Digepath, a GI-specialized foundation model using a dual-phase pretraining framework (Fig. 1). In Phase I, a ViT-based encoder pretrained on 353,478,334 multi-scale images (2.5×, 5×, 10×, and 20×) from 210,043 WSIs using self-supervised learning approach to capture gastric domain-specific features. Unlike current pathology foundation models pretrained on pan-tissue datasets[35–37], Digepath was pretrained on GI pathology images at varying scales, which capture domain-specific features at both fine- and coarse-grained level. Phase II introduces a region-of-interest (ROI) mining algorithm, creating a closed-loop feature-data optimization system to enhance diagnostic accuracy. Evaluated across 34 downstream tasks, Digepath achieved state-of-the-art performance on 33 benchmarks (Fig. 1c), demonstrating superior capabilities in pathological diagnosis, protein expression status prediction, genetic mutation prediction, prognostic assessment, and magnification-invariant tissue classification compared to existing foundation models. The translated early cancer screening module displayed robust performance across nine medical centers, achieving an average sensitivity of 99.70% coupled with 89.30% specificity. Furthermore, we introduced DigeTools—an end-to-end agent pipeline that integrated automated feature extraction, cancer detection, subtype identification, and interactive reporting. This work establishes a new paradigm for pathology-specialized AI.

## Results

### 1. Dual-phase pretraining

We developed a two-stage self-supervised framework for GI pathology analysis. First, a ViT-L model (Digepath-V1) was pretrained using DINOv2[47] on 210,043 WSIs (Fig. 1a and Supplemenary Table 1). Next, expert pathologists (with more than 10 years of clinical experience) annotated 471,443 diagnostic regions (2,048 × 2,048 pixels at 0.42 μm/pixel, Supplemenary Table 2, 3) from 26,320 WSIs to train a tumor classifier. This classifier processed the original dataset to identify 1,305,328 tumor regions ( Supplemenary Table 4), subdivided into 31,327,872 million patches with size of 256 × 256. An equal number of non-tumor patches were randomly sampled to create a multi-scale dataset including 83,206,828 patches for fine-tuning, yielding the enhanced Digepath-V2 model.

In four diagnostic tasks, it outperformed Digepath-V1 by an average balanced accuracy (ACC) improvement of 3.81% (ESO-AS: 3.58%, ESO-2cls: 3.43%, R-X:

1.51, and LHN-3cls: 6.73%), as demonstrated in Fig. 2a and Supplementary Table 5–**7**. Digepath-V2 demonstrated an average improvement of 4.18% over Digepath-V1 in TNM staging tasks (4.78% for gastric TNM and 3.58% for intestinal TNM). For PD-L1 expression status prediction (positive: CPS≥1; negative: CPS<1, see methods), Digepath-V2 achieved a 3.00% higher ACC than Digepath-V1, while the improvement was 0.49% for microsatellite instability (MSI) status prediction. These results validate the effectiveness of two-stage pretraining.

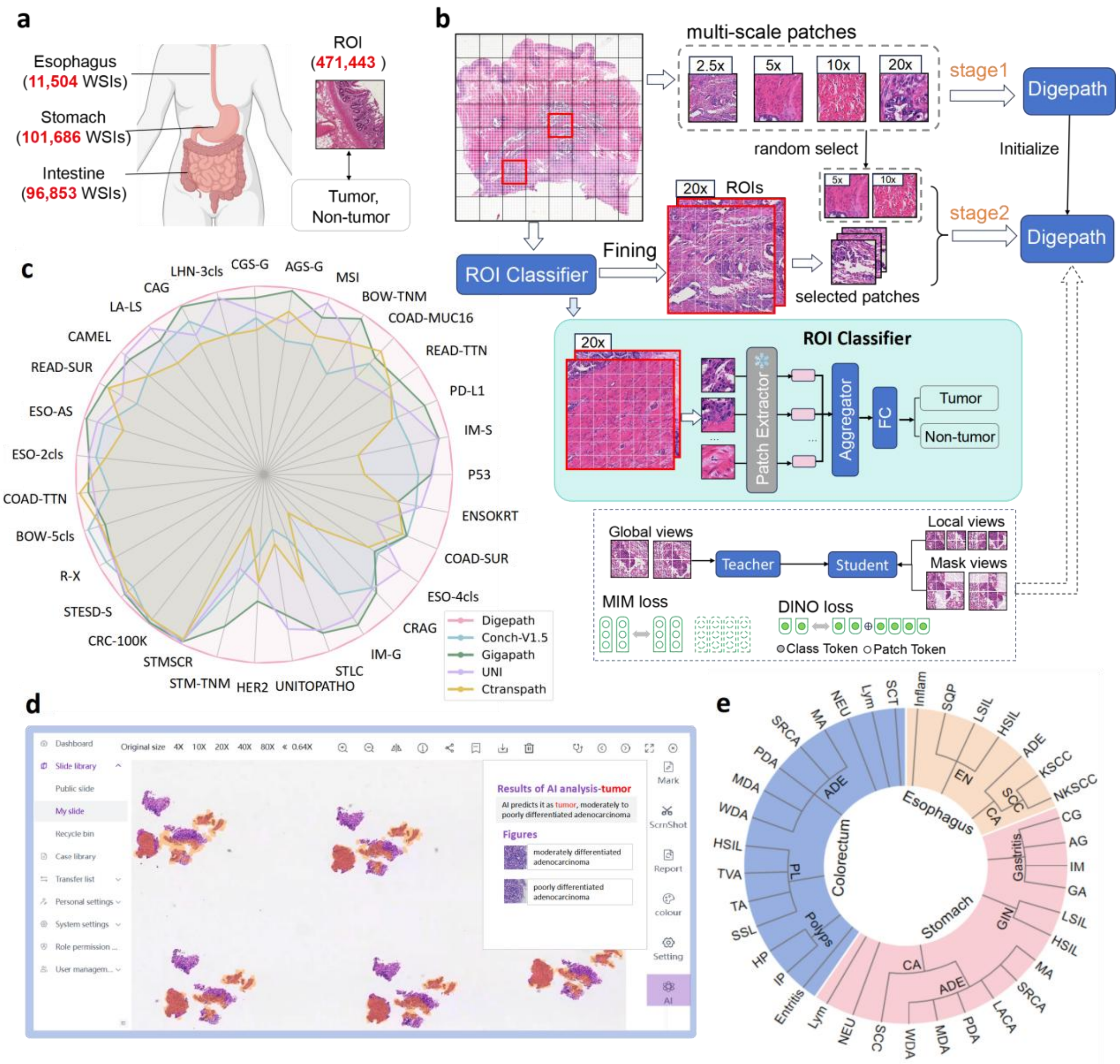

**Figure 1 | Overview of Digepath.** Digepath is a visual foundation model pretrained in two stages via the DINO-V2 framework on 210,043 GI pathology WSIs. **a.** Distribution of GI datasets included in Digepath pretraining. **b.** Two-stage pretraining scheme: In Phase I, a ViT-based encoder was pretrained on multi-scale (2.5×, 5×, 10×, and 20×) WSIs using self-supervised learning approach to capture gastric

domain-specific features. Phase II developed an ROI mining algorithm to fully utilize diagnostically valuable areas in WSIs, establishing a closed-loop enhancement mechanism of feature optimization-data refinement (see Methods). **c.** Digepath demonstrates state-of-the-art performance across a comprehensive benchmark of 33 downstream GI pathology tasks. **d.** Engineering implementation of the early-cancer screening module. **e.** Spectrum of clinical diagnoses in GI pathology associated with downstream tasks.

## 2. Multiscale pretraining enables robust representation

Unlike conventional pathology models limited to single magnifications, Digepath introduces a clinically inspired multiscale pretraining framework spanning four diagnostic resolutions (2.5×, 5×, 10×, and 20×). This design overcomes a fundamental limitation in digital pathology, where standard 224 × 224 pixels evaluation protocols compromise morphological interpretation of critical features like nuclear atypia and tissue architecture. Systematic evaluation for the classification of gastric epithelial tumors and hyperplastic lesions (STLC, see Methods), which is the most anatomically complex domain with 11 distinct subtypes across 224 × 224 to 1,120 × 1,120 pixels revealed Digepath's unique scale adaptability. Performance peaked at 672 × 672 (ACC: 95.46%) with only 0.91% variation across 5 times scale changes (Gigapath: 2.02%), ultimately achieving an ACC of 96.31% through integrated multiscale predictions, as demonstrated in Fig. 2b and Supplementary Table 8–13.

In STLC, downsampling to 224 × 224 obscured critical fine-grained details of high-grade intraepithelial neoplasia (HIN), such as the enlarged nuclei, coarse chromatin, and loss of cellular polarity, that remained detectable by Digepath at low resolutions (Fig. 2f). These observations demonstrate that Digepath encodes semantically meaningful representations that are largely invariant to image resolution, a capability of matching pathologists' multi-scale diagnostic workflow.

## 3. Routine clinicopathological diagnosis

Aligned with the WHO Classification of Digestive System Tumours, this study established a comprehensive validation framework spanning three anatomical regions (esophageal, gastric, and intestinal) across 24 clinical tasks, incorporating classification and segmentation at both ROI and WSI-level. Digepath demonstrated superior diagnostic performance across all evaluation metrics compared with well-established foundation models (Fig. 2, 3, Extended Data Fig. 1–4, and Supplementary Table 14–37).

**3.1 Stomach**

We used multiple instance learning (MIL[48]) method for automated pathological grading of non-neoplastic GI lesions in accordance with WHO diagnostic criteria. Digepath demonstrated good performance in ACC across three critical diagnostic categories (Fig. 2c and Supplementary Table 14–16): grading assessment of chronic gastritis (CGS-G; 94.67% ), acute inflammatory activity (AGS-G; 88.31%), and intestinal metaplasia (IM-G; 76.44%). Additionally, we evaluated gastric epithelial atrophy based on histomorphology and achieved an ACC of 86.06% (CAG), as outlined in Supplementary Table 17. Comparative results are provided in Fig. 2c. Fig. 2e demonstrates that during chronic gastritis assessment, the model focuses on lymphocyte- and plasma cell-enriched regions, whereas for acute activity grading, it targets neutrophil-enriched areas, which is aligning with the clinical pathological diagnostic rationale.

As evidenced in the "Multiscale pretraining enables robust representation" section and Supplementary Table 18, DigPath demonstrated superior performance for the classification of gastric epithelial tumors and hyperplastic lesions (STLC) across multiple scales. T-SNE visiualization and attention heatmaps of ROI are shown in Fig. 2d and Extended Data Fig. 1a–c. These results highlight Digepath's dual capability of diagnosis in both non-neoplastic and neoplastic lesions.

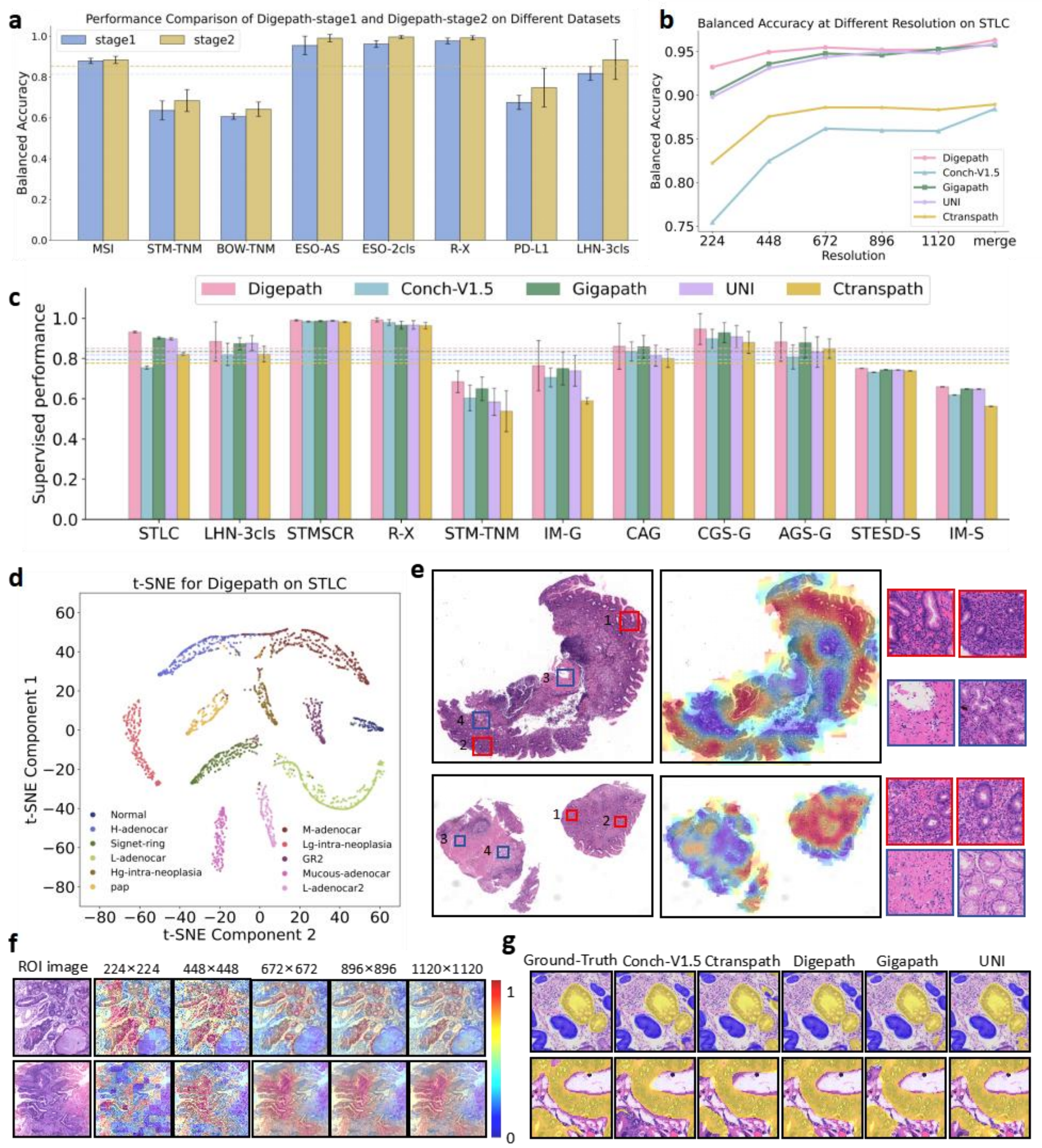

**Figure 2 | Two-stage pretraining of Digepath and its performance on gastric diagnosis. a.** Comparison of Digepath performance after the first and second pretraining stages. **b.** Multi-resolution evaluation on STLC ( 3,435 ROIs) across various methods. **c.** Comparative performance of different models on STLC. The balanced accuracy metric (y-axis) was used to evaluate the performance of Digepath on datasets of STLC (3,435 ROIs), LHN-3cls (92 WSIs), STMSCR (47,729 ROIs), R-X (77 WSIs), STM-TNM (60 WSIs), IM-G (134 WSIs), CAG (115 WSIs), CGS-G (100WSIs), and AGS-G (69 WSIs), while mean intersection over union (mean IoU) served as the evaluation metric (y-axis) for STESD-S (890 ROIs), and IM-S (227

ROIs). **d.** T-SNE visualization of Digepath features on STLC. **e.** Attention heatmaps for gastritis and acute activity grading tasks. **f.** Attention heatmaps of Digepath on STLC. **g.** Visualizations of Digepath outputs on intestinalized/non-intestinalized gland segmentation and ESD tumor region segmentation tasks.

### 3.2 Intestine

On the CRC-100K 10-class colorectal tissue classification task, Digepath achieved state-of-the-art ACC of 95.24% (Fig. 3a and Supplementary Table 19), outperforming the previous best method (Gigapath: 94.97%) by 0.27%. For intestinal polyp classification evaluated on the UNITOPATHO dataset, Digepath attained 85.92% ACC (Fig. 3a and Supplementary Table 20), representing a significant 4.45% improvement over UNI. The advantage persisted in adenoma identification (CAMEL dataset), where our model achieved an ACC of 92.36% (UNI: 91.48%; Fig. 3a and Supplementary Table 21). Notably, in the clinically critical classification of colorectal epithelial tumors and hyperplastic lesions (BOW-5cls), Digepath maintained robust performance (80.73%), consistently surpassing all competing methods (Fig. 3a and Supplementary Table 22). These results collectively establish Digepath as a new benchmark in computational pathology for colorectal tissue analysis. Attention heatmaps of intestinal adenomas in Fig. 3e also demonstrates that Digepath encodes semantically meaningful representations that are invariant to image resolution.

### 3.3 Esophagus

Digepath also demonstrated excellent diagnostic capability in the analysis of esophageal pathologies (Fig. 3a, Fig. 3c, and Supplementary Table 24–26). In the 4-class classification of epithelial tumors and hyperplastic lesions (ESO-4cls) at WSI-level, Digepath achieved 80.41% ACC, representing a substantial 2.88% improvement over the second-best approach (UNI: 77.53%).

When evaluated on another common clinical diagnostic task of distinguishing between keratinizing and non-keratinizing squamous tumors (ENSOKRT), Digepath achieved an ACC of 78.75% (Fig. 3a), exceeding the nearest competitor (UNI:

75.18%) by 3.57%. Fig. 3c illustrates that the model focuses on keratin pearls when predicting keratinizing squamous cell carcinoma (SCC).

### 3.4 Early cancer screening

In early cancer screening tasks across three major anatomical sites (stomach: STMSCR, colorectum: BOWSCR, and esophagus: ESO-2cls), Digepath demonstrated superior performance despite high baseline ACC among all evaluated models. For STMSCR using ROI, Digepath achieved an ACC of 99.01% (Fig. 2c), surpassing the second-best model by 0.23% (UNI, 98.78%). Similarly, on BOWSCR (ROI-based), it attained an ACC of 99.78% (Fig. 3a), exceeding the nearest competitor by 0.17% (UNI, 99.61%). Notably, on ESO-2cls, which is processed in WSI, Digepath achieved an ACC of 99.63% (Fig. 3b), outperforming the runner-up by a significant margin of 1.96% (Gigapath, 97.67%). More details could be available in Extended Data Fig. 2 and Supplementary Table 27–29.

### 3.5 Segmentation task

Accurate tumor segmentation serves as a fundamental pillar of modern computational pathology, enabling quantitative histopathological analysis. Our study advances this field through three clinically relevant benchmarks. First, we present an enhanced TransUnet framework[49] incorporating novel encoder architectures with pathological foundation models. When evaluated on the CRAG dataset for colorectal mucosal gland segmentation, the Digepath encoder achieved state-of-the-art performance (IoU: Digepath vs. Gigapath = 82.21% vs 79.82%; Fig. 3a, Extended Data Fig. 3c, and Supplementary Table 30). To enable quantitative metaplasia grading, we developed specialized segmentation for intestinal metaplastic and non-metaplastic glands (IM-S). Digepath attained IoU of 70.37% (surpassing Gigapath by 0.77%; Fig. 2c, Extended Data Fig. 3a, and Supplementary Table 31). Moreover, to address the unmet need for precise tumor margin delineation in ESD, we curated a clinically-annotated dataset of 4,455 patches (STESD-S). As a result, Digepath achieved a mean IoU of 85.42% (Gigapath: 84.94%; Fig. 2c, Extended Data Fig. 3b, and Supplementary Table 32).

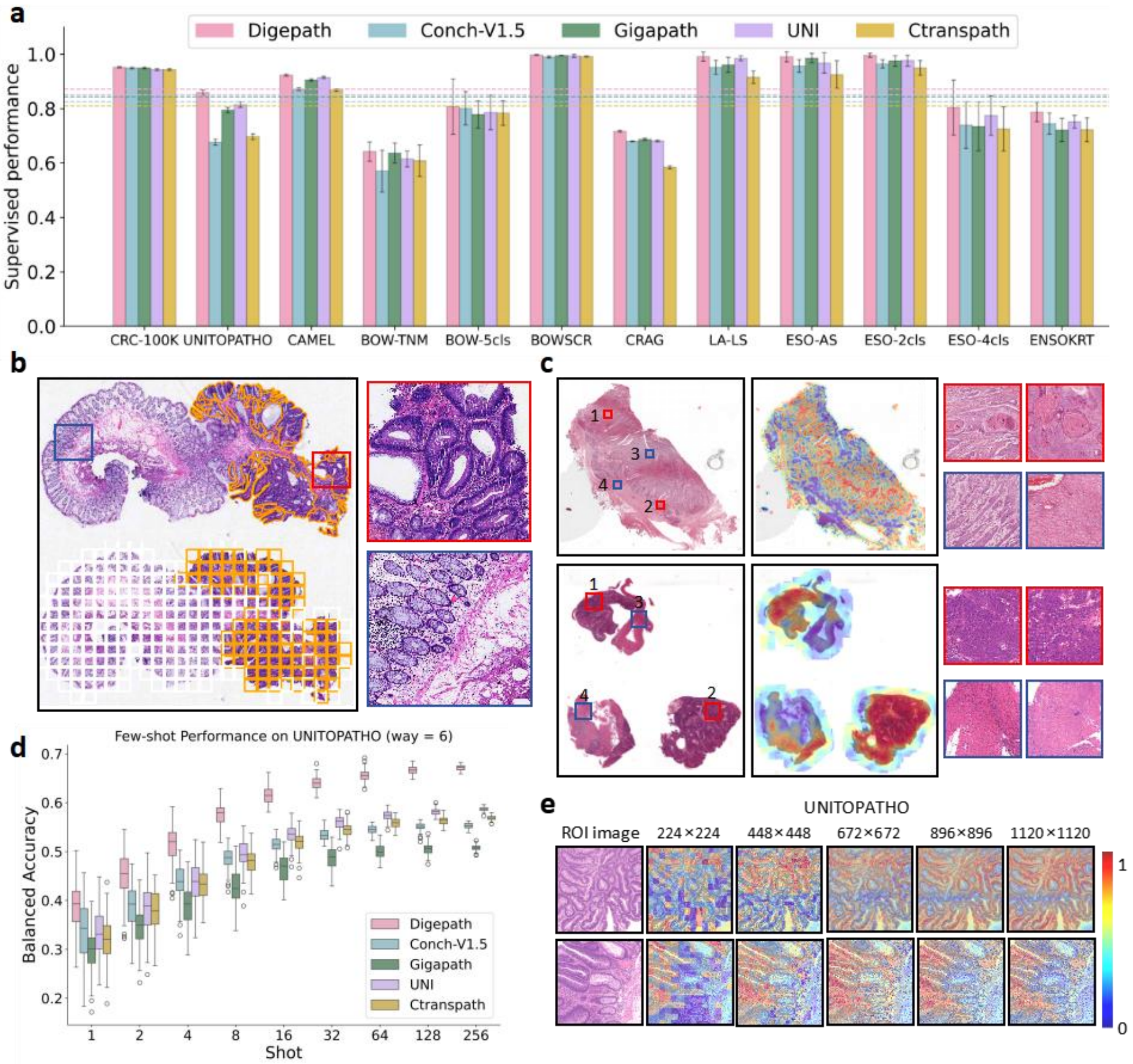

**Figure 3 | Performance of Digepath on intestinal and esophageal diagnostic tasks.** **a.** Comparative evaluation of different models on 12 intestinal and esophageal pathology tasks. Evaluation metrics (y axis) include balanced accuracy for datasets of CRC-100K (7,180 ROIs), UNITOPATHO (2,601 ROIs), CAMEL (4,621 ROIs), BOW-TNM (199 WSIs), BOW-5cls (337 WSIs), BOWSCR (30,063 ROIs), LA-LS (77WSIs), ESO-AS, ESO-2cls (172 WSIs), ESO-4cls (172 WSIs), ENSOKRT (67 WSIs), and mean IoU specifically for CRAG (40 ROIs). **b.** ROI-based early cancer screening of intestine: pathologist-annotated tumor regions (top left) vs model predictions (bottom left); orange boxes denote predicted tumor regions and white boxes denote predicted non-tumor regions. Insets (right) correspond to the red and blue boxes in the top-left panel, showing pathologist-annotated tumor and non-tumor regions. **c.** Attention-based visualizations for esophageal carcinoma prediction: top

row shows keratinizing SCC prediction; bottom row shows visualizations for esophageal carcinoma prediction. **d.** Few- shot performance comparison (K = 1, 2, 4, 8, 16, 32, 64, 128, 256) of multiple models on UNITOPATHO. **e.** Attention heatmaps of Digepath at various resolutions in the intestinal classification task.

### 3.6 TNM staging

We evaluated pathological staging performance using 1–3 representative tumor sections from surgically resected gastric and intestinal specimens. The American Joint Committee on Cancer (AJCC) staging system, which incorporates tumor histotype, invasion depth, lymph node involvement, and distant metastasis, served as our reference standard[50,51]. Notably, our approach relied solely on H&E-stained tumor sections for direct staging prediction, without ancillary clinical or imaging data. Digepath demonstrated superior staging ACC compared to competing models: 68.46% versus 64.97% (Gigapath) for gastric cancer (STM-TNM; Fig. 2c and Supplementary Table 33) and 64.24% versus 63.68% for intestinal cancer (BOW-TNM; Fig. 3a and Supplementary Table 34). Extended Data Fig. 4c indicates that when predicting stage IV colorectal cancer, Digepath primarily focuses on regions exhibiting full-thickness tumor invasion through the bowel wall. While Extended Data Fig. 4d reveals its attention to tumor-infiltrated mucosal layer and muscularis propria for stage II gastric cancer prediction, disregarding uninvolved areas.

### 3.7 Challenging pathological diagnoses

The histopathological distinction between poorly differentiated SCC and poorly differentiated adenocarcinoma in upper GI specimens poses particular diagnostic difficulties as they progressively lose their defining morphological characteristics. This morphological ambiguity routinely necessitates ancillary immunohistochemical studies for definitive classification in clinical practice. Notably, Digepath demonstrated exceptional diagnostic capability solely based on H&E-stained sections (LA-LS). The algorithm achieved ACC of 99.16% in discriminating these challenging subtypes, with 0.64% improvement over existing methods (Fig. 3a and Supplementary Table 35). When identifying poorly differentiated adenocarcinoma, Digepath focuses as much as possible on the cancerous areas that still retain minimal glandular structures (Extended Data Fig. 4a). While diagnosing SCC, it prioritizes those solid tumor nests with sheet-like or clustered patterns and densely stained boundaries (Extended Data Fig. 4b).

Histopathological differentiation among reactive hyperplasia, low-grade intraepithelial neoplasia (LIN), and HIN in gastric biopsies remains a significant diagnostic challenge. We analyzed 384 gastric biopsy slides from four medical centers (LHN-3cls). Our diagnostic model demonstrated an ACC of 88.52% in this challenging task, outperforming the next-best method by 0.88% (Fig. 2c,and Supplementary Table 36).

Distinguishing xanthoma from signet-ring cell carcinoma in GI biopsies poses diagnostic difficulties, especially among junior pathologists. We curated a dataset of 400 slides including xanthoma and signet-ring cell carcinoma (R-X). The model achieved a near-perfect discrimination (ACC: 99.22%), representing 1.33% improvement over the second-best method (Conch-V1.5: 97.89%, Fig. 2c and Supplementary Table 37).

## 4. Molecular profiling and prognostic prediction outcomes

Tumor molecular profiling guides therapeutic decision-making and is indispensable for precision oncology. We present a comprehensive evaluation of Digepath's ability to infer molecular profiling directly from histopathology images, which are prediction of therapeutic protein target expression, determination of MSI status, and genetic mutations in GI malignancies.

### 4.1 Prediction of therapeutic protein target expression

Digepath demonstrated robust predictive performance for three therapeutically relevant protein biomarkers in gastrointestinal oncology, achieving ACC >0.7 for all targets (Fig. 4a and Supplementary Table 38–40). For PD-L1 expression prediction, the model achieved an ACC of 74.83%, representing a 1.69% improvement over Gigapath (AUROC = 73.14%). In HER2 amplification detection, Digepath showed strong predictive value (AUROC = 83.27%), outperforming Gigapath by 8.20%. The system also exhibited high diagnostic AUROC for P53 mutation-type identification (AUROC = 72.23%), surpassing Gigapath's performance by 2.40%.

### 4.2 Prediction of MSI status

MSI represents a well-established biomarker for predicting response to immune checkpoint blockade therapy in colorectal cancer. Digepath achieved an AUROC of 88.41% on the self-built dataset, exceeding UNI (87.58%) by 0.83% (Fig. 4a). Attention analysis revealed that regions containing solid tumor components, luminal necrosis, and tumor-infiltrating lymphocytes received high model attention (Fig. 4e and Supplementary Table 41).

### 4.3 Prediction of recurrent genetic alterations in GI cancers

We conducted an analysis of genetic mutations with histopathological images, which exhibited some morphological signals associated with *MUC16* and *TTN* mutations on TCGA-COAD, and *NRAS* mutation on TCGA-READ. Among four prediction tasks, three achieved AUROC values exceeding 60% (Fig. 4d and Supplementary Table 42–45). achieving the best AUROCs of 65.61%, 62.37%, and 60.55%, respectively.

### 4.4 Prognostic prediction in GI oncology

We also implemented an augmented Digepath architecture to generate histomorphology-based survival models with publicly accessible cohorts. Digepath achieved statistically robust discrimination between favorable (long-term) and poor (short-term) survival subgroups for TCGA-COAD, significantly outperforming existing approaches with a concordance index of 71.82% (Fig. 4b, Fig. 4c, and Supplementary Table 46), representing relative improvements of 3.80% over Conch-V1.5 (68.02%). The model's superior precision was further evidenced by significantly tighter confidence intervals (CIs) in Kaplan-Meier analyses (P<0.01; Extended Data Fig. 5), indicating enhanced prognostic reliability.

Extended Data Fig. 6b shows DigPath assigned higher attention weights to regions displaying dense lymphocytic infiltration and preserved tissue architecture with mild atypia in favorable-prognosis patients, while Fig. 6a highlighted regions lacking immune infiltration and exhibiting poorly differentiated tumor morphology in poor-prognosis cases. Other comparative models primarily focus on non-tumor regions.

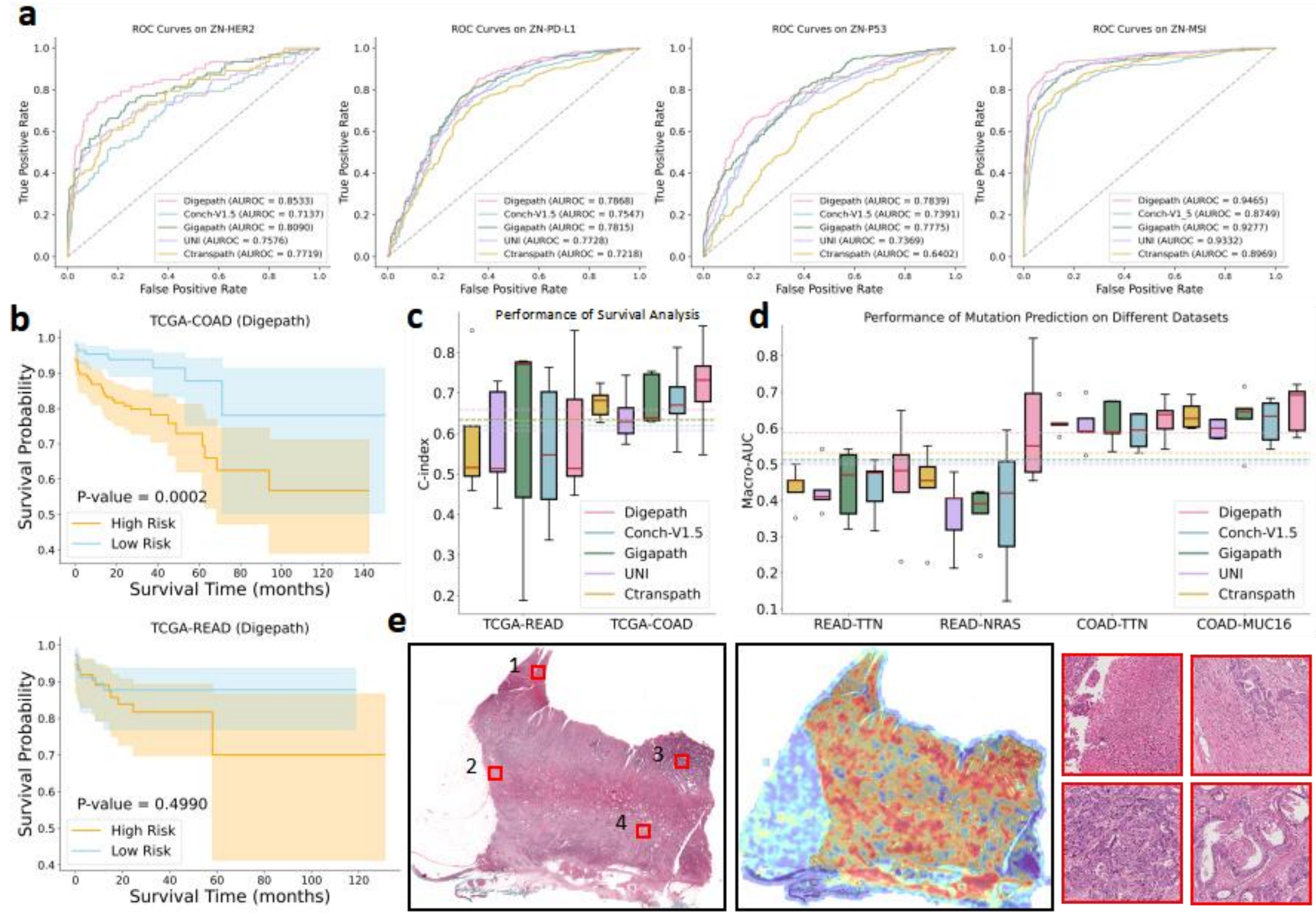

**Figure 4 | Molecular and prognostic prediction using Digepath. a.** Molecular marker prediction performance (HER2, PD-L1, P53, MSI) across four in-house datasets (80 WSIs, 151 WSIs, 142 WSIs, and 194 WSIs ). **b.** Kaplan–Meier survival analysis predictions on TCGA cohorts: TCGA-COAD (top, 82 WSI) and TCGA-READ (bottom, 31 WSIs). **c.** Comparison of concordance-index (C-index) for survival prediction across methods on TCGA-COAD and TCGA-READ. **d.** Comparative mutations prediction of *TNN*, *NARAS*, and *MUC16* on TCGA-COAD and TCGA-READ ( 81 WSIs and 26WSIs). **e.** Attention heatmaps of Digepath for MSI status prediction.

# 5. Clinical translation and implementation

## 5.1 Early cancer screening

Pathologists face the critical yet challenging task of detecting rare early-stage malignancies among vast numbers of gastrointestinal biopsy specimens, which is a time-intensive process. To transform this paradigm, we implemented an AI-powered early gastric cancer screening module based on Digepath and validated it through a multi-center study across nine Chinese medical centers selected for geographic and institutional diversity (Fig. 5h). Following WHO 5th edition criteria, we classified LIN, HIN and malignancies as positive (657 WSIs), benign polyps and chronic gastritis as negative (10,567 WSIs). The module achieved an average ACC of 89.99%, with a sensitivity of 99.70% and specificity of 89.30%. Notably, it reached perfect sensitivity at seven participating hospitals and exceeded 90% specificity at five institutions. Detailed site-specific metrics are shown in the Fig. 5g and Supplementary Table 47, 48.

Across these hospitals, the module successfully identified one neuroendocrine tumor (NET), one signet ring cell carcinoma (SRCC), one highly differentiated adenocarcinoma, one poorly differentiated carcinoma, and six cases of LGIN. The NET case, classified as G1, exhibited mild cytological features and was easily overlooked. The SRCC case involved a small focus located at the edge of the biopsy specimen within an inflammatory background, which was similarly prone to misdiagnosis. Nevertheless, the model has correctly flagged the lesion, which was subsequently confirmed via immunohistochemical staining for CEA and CK. In addition, the model accurately identified four more SRCC cases that were not missed by pathologists but posed diagnostic challenges due to their morphological resemblance to histiocytes or because they consisted of only a few scattered signet ring cells. The highly differentiated adenocarcinoma case displayed features closely resembling normal gastric epithelium, and its small biopsy volume further complicated diagnosis. However, the model correctly localized the subtle serrated structures indicative of malignancy. One LGIN case that was initially missed by the model was later confirmed as positive via ESD resection. Furthermore, three cases initially diagnosed as LGIN were reclassified as intestinal metaplasia after expert consensus review, aligning with the model's original prediction of non-neoplastic. The model's two missed cases included one acid-secreting adenoma and one additional instance of LGIN. Details could be available in Fig. 5a–f.

### 5.2 Digestive pathology agent system enabling pathology report generation

We also developed an end-to-end pipeline of agent system for digestive pathology (Extended Data Fig. 9e). Taking a WSI together with a user text prompt as inputs, the system leverages the DigeTools library to sequentially perform cancer detection, subtype identification, and ROI-level report generation through multi-turn dialogue. First, the agent activates the Feature Extraction module, partitioning the WSI into patches with the size of 256 × 256 at 20× magnification. A pretrained Digepath encoder generates embeddings of these tiles, which are immediately analyzed by the tumor detection module for early cancer screening. The system then proceeds through sequential diagnostic modules, ultimately identifying the case as non-keratinizing SCC while automatically generating detailed cytologic and histologic descriptions through a large language model (LLM). For local analysis, the system performs ROI Selection across the entire slide, using the ROI Finder pinpointing images with high attention scores. Finally, the DigeCaption module produces a comprehensive diagnostic report combining quantitative data with qualitative interpretation, completing an integrated workflow from macroscopic detection to microscopic analysis.

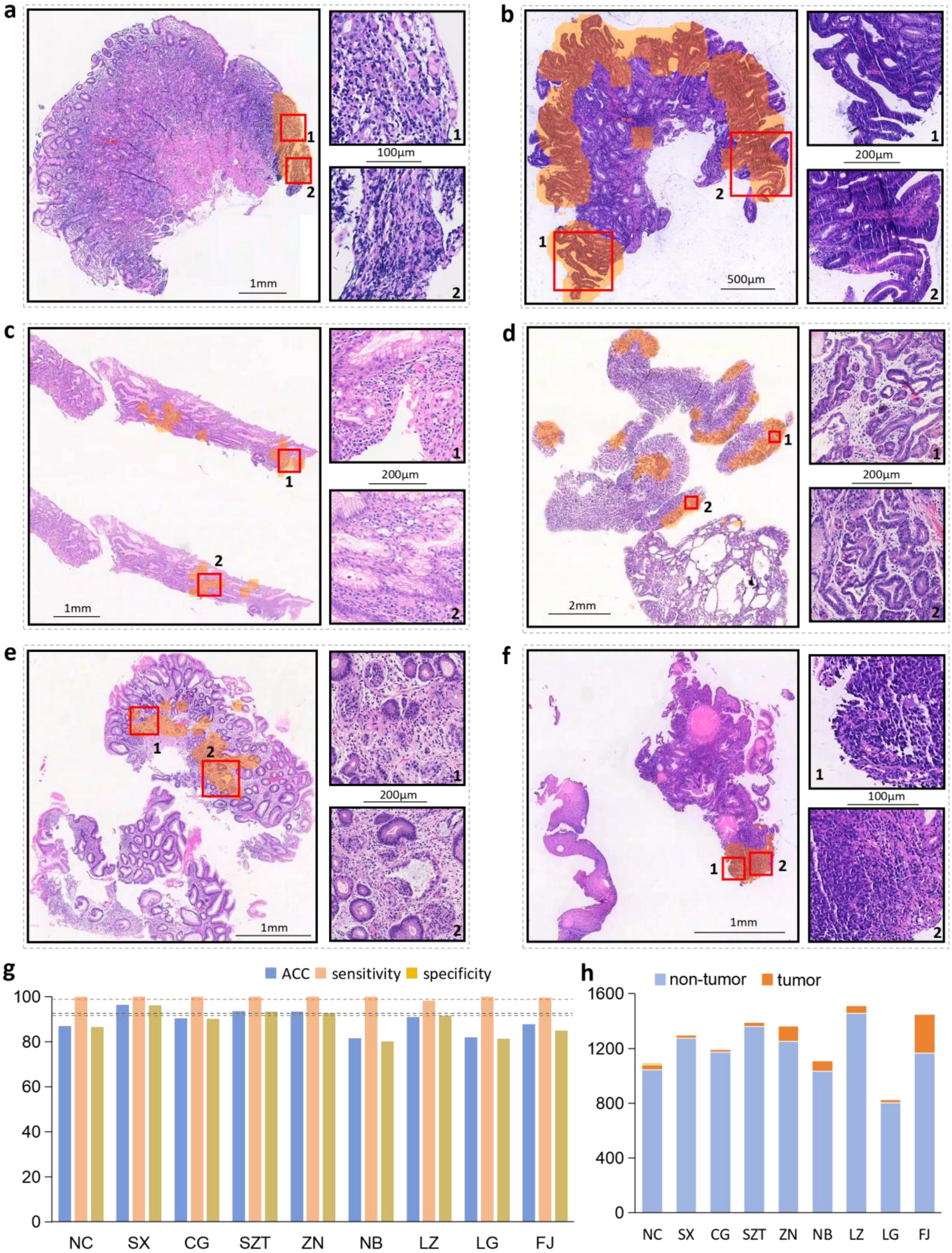

**Figure 5 | Early-cancer screening results. a–f.** Cases where the model assisted pathologists in identifying previously missed diagnoses. **a.** signet-ring cell carcinoma. **b.** low-grade epithelial neoplasia. **c.** highly differentiated adenocarcinoma. **d.** fundic gland tumor. **e.** neuroendocrine tumor. **f.** poorly differentiated carcinoma. **g.** Performance of the early cancer screening module across nine independent centers. **h.** Data distribution from the nine medical centers.

## 6. Few-shot learning

We assessed Digepath's label efficiency on the datasets of STLC and UNITOPATHO using the non-parametric SimpleShot framework—a robust baseline widely adopted in the few-shot classification literature[52]. In few-shot learning, the choice of 'way' has a significant impact on task difficulty and model performance. Typically, increasing the number of ways indicates a greater number of classes to classify, thereby raising the task difficulty. Detailed experimental protocols and performance results are provided in the Methods and Extended Data Fig. 7.

Across different tasks and ways evaluation, we found that Digepath is a powerful few-shot learner with markedly higher label efficiency than other pretrained encoders. When comparing median performance at varying sample sizes, Digepath consistently surpasses the next-best encoder at every shot across two tasks (Fig. 3d and Extended Data Fig. 7). Remarkably, Digepath's 8-shot performance in both tasks can exceed the maximum performance achieved by other encoders over 1,000 trials. Meanwhile, as the number of the way increases, the gap in few-shot performance between Digepath and other models becomes increasingly evident. This demonstrates that Digepath can better leverage its superior capabilities of label efficiency and representation quality in more challenging gastrointestinal tract classification tasks.

## 7. ROI retrieval

ROI retrieval is implemented through Prototypical Network[53] (ProtoNet). The ProtoNet first convert all training images into embedding vectors, then performs mean-pooling on embeddings of the same category to obtain prototype representations. We benchmarked histopathological image retrieval across two ROI-level tasks, with each test sample classified based on its highest similarity to class prototypes. Detailed experimental procedures and results are provided in the Methods, Extended Data Fig. 8, and Supplementary Table 49, 50.

On two retrieval tasks, Digepath consistently outperformed competing encoders, demonstrating superior retrieval ACC across diverse settings. On STLC dataset,

Digepath achieved a 11.11% gain over the next best encoder (Digepath: 74.07%, UNI: 62.96%). On UNITOPATHO dataset, the performance gap narrowed—Digepath exceeded UNI by 8.26% (Digepath: 67.21% vs UNI: 58.95%) —likely reflecting the pronounced morphological distinctions.

For both papillary carcinoma and HIN categories in the STLC dataset, we selected the top five images most similar to each model's prototype. Results demonstrated that Digepath accurately identified representative images for both categories (Extended Data Fig. 8), confirming its superior feature representation capability. This finding was consistently replicated in the UNITOPATHO dataset.

## Discussion

This study represents a transformative advancement in computational pathology for GI disease, with three fundamental innovations that address critical challenges in the field. Firstly, we introduce the concept of specialty-specific foundation models for pathology. Current pathological foundation models face limitations in domain-specific diagnostic performance. Our GI-specialized foundation model resolves the generalizability-specialization trade-off. Pretraining on GI datasets, Digpath maintains transfer learning benefits while enabling key diagnostic capabilities including early cancer detection (>99%), SCC and adenocarcinoma differentiation (99.16%), and xanthoma versus signet-ring cell carcinoma discrimination (99.22%), advancing clinically deployable AI in pathology. Secondly, we develop a novel two-stage progressive training architecture with methodological breakthroughs. The first stage employs multi-resolution image analysis combined with DINOv2 self-supervised learning on 353,478,334 patches of the digestive tract to extract universal features. The subsequent stage implements dynamic ROI selection with contrastive learning optimization, driving significant performance improvements across 33 downstream diagnostic tasks in gastrointestinal pathology. Thirdly, we established a closed-loop framework for clinical translation that seamlessly integrates clinical needs, technological development, and real-world application.

Furthermore, we built a comprehensive validation framework, covering 21 morphological diagnostic tasks, three segmentation tasks, eight molecular profiling prediction tasks, and two survival prediction tasks. From an engineering perspective, a gastric biopsy early cancer screening module was developed and deployed in routine testing across multiple medical centers. These methodological and translational innovations effectively bridge the critical gap between computational pathology research and clinical practice. Our system shows particular promise for enhancing early cancer detection in resource-limited settings.

While attention-based MIL (ABMIL[48]) framework demonstrates robust performance in WSI-level prediction tasks such as non-neoplastic lesion grading and tumor histological subtyping, it has inherent limitations in modeling the complex spatial relationships between tissue patches across entire slides, which is a critical component of comprehensive pathological assessment that requires integration of both local morphological features and global architectural patterns. Current computational approaches including graph neural networks and vision transformers present promising solutions to these limitations through their ability to explicitly encode spatial dependencies between distant tissue regions.

Clinically, our deployment platform is designed for continuous evolution through adaptive features that incorporate new diagnostic modules like inflammatory bowel disease activity scoring while maintaining rigorous validation standards. The system's real-time optimization using hospital-derived data and dynamic updating protocols will facilitate its development into a more reliable diagnostic assistant that remains aligned with evolving clinical requirements.

**Methods**

In recent years, foundation models have demonstrated remarkable transfer capabilities in natural image analysis. Architectures such as Vision Transformer (ViT), known for their robust global feature embedding capability, have been widely applied across diverse tasks[30]. The rapid advancement of self-supervised learning methods (e.g., DINOv2[47] and MoCov3[54]) has further optimized feature representation, significantly enhancing model performance in varied scenarios. In the field of computational pathology, researchers have developed domain-specific foundation models, such as UNI, Gigapath, and TITAN[36–38]. These models leverage the advantages of self-supervised learning to construct generalized feature representations by pretraining on large-scale pathological datasets.

Based on the DINOv2 framework, we propose a two-stage pretrained model tailored for gastrointestinal diseases. It fully utilizes ROI information and disease-specific characteristics to improve diagnostic accuracy for gastrointestinal pathology.

**Dual-phase pretraining for the gastrointestinal pathological foundation model**

**Pretraining on multi-scale gastrointestinal pathological images.** The overall framework is illustrated in Fig. 1b. Distinct diagnostic tasks necessitate examination at specific magnification, for instance, 2.5× and 5× for macroscopic tissue assessment, 10× for analyzing cellular morphology, and 20× for assessing nuclear atypia. To address these multi-scale diagnostic requirements, we pretrained the model with multi-magnification (2.5×, 5×, 10×, and 20×) images as inputs. This approach captures comprehensive pathological information spanning macroscopic tissue morphology to microscopic cellular features, thereby establishing a comprehensive multi-scale database for gastrointestinal pathology images.

**Visual foundation models pretraining based on DINOV2.** This study employs the DINOv2 framework for visual pretraining to enhance feature extraction and domain-specific semantic understanding for gastrointestinal pathology image analysis. DINOv2 advances self-supervised learning through teacher-student distillation tailored for ViTs. By integrating self-distillation with masked image modeling, it learns transferable representations without manual annotations. Key innovations include enhanced augmentations and dual-loss optimization, both of which significantly improve its self-supervised learning capacity. Details of training is available in Supplementary Table 51.

**Dual-phase ROI-based optimization framework.** Our train-refine-repeat framework tackles the needle-in-a-haystack problem of finding rare diagnostic regions in gigapixel WSIs by: (1) building generalized feature representations through large-scale pretraining, then (2) using these features to identify and prioritize diagnostically critical regions for focused learning. This bidirectional optimization elevates both data quality and model performance through successive iterations.

we built a fine dataset containing 471,443 histopathologically confirmed annotations. The manually annotated ROIs are used as training data for a MIL-based classifier[55–58], facilitating automated screening in subsequent data curation. By learning the discriminative contributions of individual instances, the model automatically assigns instance-specific weights, thereby adaptively focusing on the most classification-relevant regions. Details of ROI classifier is outlined in Supplementary Table 52.

**Dynamic ROI selection strategy.** Following the training of the binary ROI classifier, we designed a confidence-weighted adaptive sampling algorithm to automatically refine the training dataset. The trained classifier was applied to the full first-stage dataset for inference, and ROI selection was guided by predicted probabilities. For WSIs containing at least one predicted "tumor" ROI, we selected the top $N_1$ ROIs with the highest classification confidence for the "tumor" class and randomly sampled $N_2$ ROIs predicted as non-tumor from the same slide. For WSIs in which all ROIs were predicted as non-tumor, we randomly selected $N_3$ ROIs. The details of the sampling strategy are defined in Equations (7)–(9).

$$N_1 = \lceil 12 \times p_{tumor}/0.7 \rceil \tag{7}$$

$$N_2 = \lceil 4 \times (1 - p_{tumor}) \rceil \tag{8}$$

$$N_3 = \text{Possion}\ (\lambda = 8) \tag{9}$$

In this sampling framework, $p_{tumor}$ denotes the predicted probability that a given ROI contains tumor tissue. Based on this strategy, we aimed to construct a refined, high-quality dataset comprising approximately 2,610,656 ROIs, with a balanced tumor to non-tumor ratio of 1:1. These selected ROIs serve as precise, task-relevant inputs for continued model training, enabling improved supervision in the second-stage pretraining process.

**Other methods and relevant parameters**

**Weakly supervised slide classification.** For WSI-level diagnostic tasks, we adopted a standard two-stage MIL framework for downstream evaluation. This pipeline begins by applying the pretrained feature extractor to all patches within a WSI, resulting in a set of encoded feature vectors in a unified embedding space. These patch-level features are then aggregated into a WSI-level representation using a gated attention-based MIL (ABMIL) architecture[48]. Depending on the specific diagnostic task, different magnification levels were employed to extract the input patches. However, for all tasks, the WSI was uniformly divided into non-overlapping patches of size 224 × 224 pixels, and the extracted patches were normalized using the same normalization parameters as those employed during pretraining of the corresponding feature extractor.

**ROI classification.** To evaluate the transferability and representational quality of pretrained features on ROI-level classification tasks, we followed the standard evaluation protocol introduced in UNI. Specifically, we employed linear probing using logistic regression with L2 regularization, where the regularization coefficient $\lambda$ was defined as100/M × C , with M denoting the feature embedding dimension and C the number of classes. The model was optimized using the L-BFGS algorithm[59] with a maximum of 1,000 iterations. For all ROI datasets, we used an input resolution of 224 × 224 pixels. For high-resolution ROI datasets, we additionally evaluated model robustness to scale variation using resolutions of 224, 448, 672, 896, and 1120 pixels, to assess the robustness of different pretrained feature extractors to input resolution changes.

**ROI attention visualization.** To further investigate the spatial attention patterns of pathology foundation models, we visualized the attention scores between the [CLS] token and patch tokens in the last Transformer layer across different input resolutions. Specifically, we conducted experiments using ROI images with resolutions of 224, 448, 672, 896, and 1120 pixels on the STLC, and UNITOPATHO datasets to examine how pretrained models attend to different regions within each ROI. As the self-supervised foundation models are trained without label supervision, the interpretability of their attention distributions with respect to class-relevant regions remains uncertain. To address this, we appended an additional Transformer layer architecturally aligned with the pretrained model at the end of the ViT backbone and fine-tuned it under supervised conditions using labeled data. We then repeated the same attention visualization procedure on the fine-tuned model, examining the attention scores between the class token and patch tokens in the new Transformer layer. The resulting maps more accurately reflected attention distributions aligned with class-specific regions, suggesting improved localization and interpretability under label supervision.

**ROI segmentation.** We conducted semantic segmentation experiments on ROI-level images using the standard TransUnet architecture[49]. TransUnet integrates a Transformer-based feature extractor into the conventional U-Net framework, effectively addressing the limitation of U-Net in modeling long-range dependencies. The Transformer encoder within TransUnet was initialized with pretrained weights from various pathology foundation models. During training, random horizontal and vertical flipping was used for data augmentation.

**Survival analysis.** The data processing paradigm for WSI-level survival prediction followed the same preprocessing pipeline as standard classification tasks, including patch extraction, feature encoding, and MIL-based aggregation. However, unlike traditional MIL classification models, which output class probabilities and final predicted labels, survival analysis models are designed to produce a risk score, a predicted survival label, and a time-dependent survival probability curve for each sample. During training, we adopted the Cox loss to optimize the model for censored survival data, which is defined as follows:

$$\mathcal{L} = -\frac{1}{N}\sum_{i=1}^{N}(E_i)\left[\theta_i - \log\left(\sum_{j:T_j \geq T_i} e^{\theta_j}\right)\right] \tag{10}$$

In equation (10), N denotes the total number of samples. $E_i$ is the event indicator for the i-th sample. $\theta_i$ represents the predicted risk score for the i-th sample. The risk set indicator matrix $R_{ij} = (T_j \geq T_i)$ defines whether sample j is at risk at the time of event occurrence in sample i.

**Few-shot learning.** For ROI-level classification cation tasks, we followed the evaluation standards established in the few-shot learning literature by adopting the SimpleShot framework[52]. In this pipeline, feature representations of C-way, K-shot samples from the support set are extracted using a pretrained feature encoder. The choice of 'way' has a significant impact on task difficulty and model performance. Typically, increasing the number of ways indicates a greater number of classes to classify, thereby raising the task difficulty. Class prototypes are then computed by averaging the normalized and centered feature vectors within each class. Predictions for the query set are obtained by computing the distance between query features and class prototypes. Each evaluation run, referred to as an episode, follows this procedure. We conducted 1,000 episodes for each dataset, using all available classes (ways). The number of shots K was varied across {1, 2, 4, 8, 16, 32, 64, 128, 256}, depending on the minimum number of available samples in each class.

**Digestive pathology agent architecture.** The current system comprises three core modules: Dige Task Suite, WSI Process, and Dige Caption. The GPT-4o engine handles natural language instruction parsing domain knowledge inference and dynamic tool orchestration. External functionalities are implemented as standardized function calls with metadata descriptors including tool summaries I/O schemas and exemplar prompts enabling context-aware retrieval and execution during reasoning processes.

Downstream Tasks unifies diagnostic models for WSI-level analysis including benign/malignant diagnosis histopathological subtyping and survival prognosis alongside ROI tasks such as classification and segmentation. Each subsystem provides an API accepting slide IDs, coordinates or feature vectors returning predictions with 95% confidence intervals to support multimodal decision fusion.

WSI Process is responsible for slide preprocessing, feature extraction, and ROI selection. Specifically, the raw WSI is first colour-normalized and then partitioned at 20 × magnification into 256 × 256-pixel tiles; each tile is mapped to a 1024-dimensional embedding through the pretrained Digepath encoder, after which the tool selected from downstream tasks computes attention weights to obtain a slide-level representation and its corresponding classification. Regions with attention scores greater than threshold $\tau$ are output as ROIs.

The Dige Caption module adopts a two-stage training strategy to enhance cross-modal descriptive capability. The backbone consists of the pretrained visual encoder Digepath and the large language model Qwen-2.5-14B-Instruct, bridged by a MLP-based projector with three layers for feature alignment. The training corpus comprises 272 k Quilt-GI image–text pairs and 18.4 k ROI-VQA samples from LZ. During stage I only the MLP-projector weights are updated, while the visual encoder and the large language model remain frozen.

**Comparisons and baselines.** To comprehensively evaluate the performance of our proposed method, we established a comparative benchmark comprising five publicly available pathology foundation models: Ctranspath[35], UNI[36], Gigapath[37], and Conch-V1.5[38].

Ctranspath was pretrained on 29,753 WSIs spanning 25 anatomical sites from the TCGA public dataset. Utilizing the MoCo-v3 self-supervised learning framework and approximately 15 million pathology tiles, it builds a Swin Transformer-Tiny–based visual encoder. UNI integrated 100,000 H&E-stained slides representing 20 tissue types and adopted the DINOv2 self-supervised paradigm to train a ViT-Large–based model on over 100 million pathology tiles, producing a general-purpose representation model. Gigapath was developed using a multi-center dataset from Providence Health in the United States, comprising 171,189 WSIs from over 30,000 patients across 28 cancer centers. Covering 31 major tissue types, this dataset enabled pretraining on 1.3 billion tiles to construct a ViT-Giant–based feature extraction system. Conch-V1.5 employs UNI as its vision tower and utilizes the native text encoder of Conch. It underwent multimodal training on a dataset of 1.26 million image-caption pairs using COCA[60]. Conch-V1.5, in conjunction with its slide encoder TITAN, demonstrated exceptional performance in tasks such as zero-shot and few-shot learning.

In all downstream task evaluations, we used the official pretrained weights provided by each of the aforementioned pathology foundation models. To ensure consistency, image normalization was performed using the mean and standard deviation parameters employed during each model's pretraining phase. For each downstream task, we maintained identical optimization hyperparameters, training steps, and model selection criteria across all models. This uniform evaluation protocol was adopted to ensure a fair and unbiased comparison of performance.

**Evaluation metrics**

For classification tasks, we use the following metrics: balanced accuracy (ACC), weighted F1-score, area under the receiver operating characteristic curve (AUCROC), sensitivity, and specificity. Balanced accuracy is equivalent to the macro-averaged recall and reflects the mean per-class accuracy. The weighted F1-score represents the class-wise F1-score averaged according to class sample proportions. AUCROC measures the area under the receiver operating characteristic curve. For semantic segmentation tasks, we use mean dice coefficient (M-Dice) and mean intersection over union (M-IoU). M-Dice calculates the macro-averaged dice score across all classes and assesses the degree of overlap between predictions and ground truth. M-IoU computes the macro-averaged intersection-over-union score across classes, reflecting segmentation precision and coverage. For survival prediction tasks, we report the concordance index (C-index), which measures the model's ability to correctly rank survival times.

**Statistical analysis**

For all semi and fully supervised experiments, we estimate 95% confidence intervals for the model performance with non-parametric bootstrapping using 1,000 bootstrap replicates. For ROI-level few-shot classification, for each C-way, K-shot setting, we randomly sample K training examples per C classes with 1,000 repeated experiments evaluated on the entire test set. For WSI-level tasks, we use 5-fold cross-validation to evaluate the performance of each model. For survival analysis tasks, we adopt the t-test to evaluate the statistical significance.

## Dataset

### Pretraining dataset

Training Dataset for the GI domain-specific foundation model in stage I was constructed in collaboration with pathology departments from three different hospitals, including Zhongnan Hospital of Wuhan University (ZN), Liuzhou People's Hospital (LZ), and Fuzhou University Affiliated Provincial Hospital (FJ). The dataset comprises a total of 210,043 WSIs scanned at a resolution of 0.25 μm/pixel, covering three major anatomical sites: esophagus (11,504), stomach (101,686), and intestine (96,853). All WSIs were anonymized to ensure compliance with privacy and ethical guidelines.

In the pretraining of stage II, three senior gastrointestinal pathologists (minimum 10 years of clinical practice) formed our validation committee assisted to construct a refined, high-quality dataset comprising 471,443 ROIs, with tumor (Low) to non-tumor ratio of 201,851:268,592. This is the largest clinically-adjudicated collection specifically designed for GI pathology AI applications.

The annotated ROI was then used to train a tumor classifier, which processed the original dataset to identify 1,305,328 tumor regions, subdivided into 31,327,872 patches with size of 256 × 256. An equal number of non-tumor patches were randomly sampled to create a multi-scale dataset including 83,206,828 patches for fine-tuning, yielding the enhanced Digepath-V2 model.

### Dataset of downstream tasks

Based on the digestive system diseases issued by WHO, we established a comprehensive benchmark comprising 34 clinically relevant tasks across three major anatomical sites.

**Early gastric cancer screening (2 classes, STMSCR)**. The dataset was collected from four medical centers, ZN, LZ, FJ, and Second Affiliated Hospital of Southern University of Science and Technology (SZT), comprising 238,643 annotated ROIs from 12,435 WSIs, each measuring 2048 × 2048 pixels at the native 20× magnification level. The dataset was divided into two classes: Class 0 included non-neoplastic conditions such as gastritis, intestinal metaplasia, reactive hyperplasia, fundic gland polyps, and hyperplastic polyps (128,575 ROIs); Class 1 included lesions such as LIN, HIN, adenocarcinoma, NET, and lymphomas (110,068 ROIs). A five-fold cross-validation protocol was employed, with each fold further split into training, validation, and test sets (167,050:23,864:47,729 ROIs). All ROI inputs were processed at 20× magnification during MIL-based classification.

**Gastric epithelial neoplasia and hyperplasia classification (11 classes, STLC).** A multi-class classification task was designed to assess model performance across a diverse set of gastric epithelial lesions. The dataset comprised 11,449 ROIs, each measuring 2,048 × 2,048 pixels at the native 20× magnification level. These ROIs were rigorously curated from four multicenter medical institutions (ZN, LZ, SZT, and FJ) and encompassed 11 distinct histopathological diagnostic categories. Class 0 included non-neoplastic lesions such as gastritis, intestinal metaplasia, reactive hyperplasia, fundic gland polyps, and hyperplastic polyps (432 ROIs). Class 1 and Class 2 corresponded to LIN (1,139 ROIs) and HIN (722 ROIs), respectively. Classes 3 to 9 included various gastric carcinomas: well-differentiated adenocarcinoma (1,273 ROIs), moderately differentiated adenocarcinoma (2,042 ROIs), poorly differentiated adenocarcinoma (1,749 ROIs), signet-ring cell carcinoma (1,085 ROIs), mucinous carcinoma (796 ROIs), other poorly cohesive carcinomas (756 ROIs), papillary adenocarcinoma (837 ROIs) and atypical hyperplasia (618 ROIs). For training and evaluation, we used train-test (8,014:3,435 ROIs) split.

**Chronic gastritis grading (3 classes, CGS-G).** This dataset, comprising 499 biopsy-WSIs, was collected from ZN and includes only biopsy samples. It is annotated for three levels of chronic gastritis: class 0 included mild chronic inflammation (171 WSIs), class 1 included moderate chronic inflammation (144 WSIs) and class 2 included severe chronic inflammation (184 WSIs). The dataset was split into training, validation, and test sets (319:80:100 WSIs), following a five-fold cross-validation protocol. All WSI inputs were processed at 20× magnification during MIL-based classification.

**Acute gastric activity grading (3 classes, AGS-G).** This dataset consisted of 348 biopsy-derived WSIs collected from ZN, annotated into three categories: class 0 comprised mild acute activity (100 WSIs), class 1 comprised moderate acute activity (121 WSIs), and class 2 comprised severe acute activity (127 WSIs). The dataset was divided into training, validation, and test sets (223:56:69 WSIs), following a five-fold cross-validation protocol. All WSI inputs were processed at 20× magnification during MIL-based classification.

**Atrophic gastritis classification (2 clasees, CAG).** This dataset comprised 571 biopsy-WSIs collected from ZN, annotated for two categories: class 0 consisted of non-atrophic (225 WSIs) and class 1 consisted of atrophic (346 WSIs). The data were split into training, validation, and test sets (364:92:115 WSIs), following a five-fold cross-validation scheme. All WSI inputs were processed at 20× magnification during MIL-based classification.

**Intestinal metaplasia grading (4 classes, IM-G).** This dataset comprised 667 biopsy-derived WSIs collected from ZN, categorized into four classes: class 0 consisted of no metaplasia (120 WSIs), class 1 consisted of mild metaplasia (216

WSIs), class 2 consisted of moderate metaplasia (64 WSIs), and class 3 consisted of severe metaplasia (267 WSIs). The dataset was split into training, validation, and test sets (426:107:134 WSIs), following a five-fold cross-validation configuration. All WSI inputs were processed at 20× magnification during MIL-based classification.

**Early colorectal cancer screening (2 classes, BOWSCR).** This dataset was constructed across three medical centers (ZN, LZ, SZT), comprising 5,837 whole-slide images (WSIs) and approximately 150,318 ROIs. The ROIs were classified into two categories: class 0 included enteritis, inflammatory polyps, and hyperplastic polyps (46,934 ROIs); class 1 included LIN, HIN, adenocarcinoma, NET, and lymphoma (103,384 ROIs). The dataset was split into training and test sets using five-fold cross-validation. Within each training fold, ROIs were further divided into training, validation, and internal test sets (105,223:15,032:30,063 ROIs). For the slide-level classification task, 20× magnification images were used as model input.

**CRC-100K tissue classification (9 classes, CRC-100K).** The CRC-100K dataset consists of 107,180 annotated regions of interest (ROIs) extracted from H&E-stained formalin-fixed paraffin-embedded (FFPE) diagnostic WSIs of 136 colorectal adenocarcinoma samples. These samples were obtained from the National Center for Tumor Diseases (NCT) tissue bank and the pathology archives of the University Medical Center Mannheim (UMM). The ROIs are labeled into nine tissue categories: adipose tissue (11,745 ROIs), background (11,413 ROIs), debris (11,851 ROIs), lymphocytes (12,191 ROIs), mucus (9,931 ROIs), smooth muscle (14,128 ROIs), normal colon mucosa (9,504 ROIs), cancer-associated stroma (10,867 ROIs), and colorectal adenocarcinoma epithelium (15,550 ROIs). For training and evaluation, we used the officially provided case-stratified training–test split (100,000:7,180 ROIs).

**UNITOPATHO colorectal polyp classification (6 classes, UNITOPATHO).** This dataset comprises 8,669 ROIs at a resolution of 1,812 × 1,812 pixels and 867 ROIs at 15,855 × 15,855 pixels, all with a spatial resolution of 0.44 μm/pixel. These ROIs were extracted and annotated from H&E-stained FFPE diagnostic WSIs of 292 colorectal polyp samples collected at the University of Turin. The ROIs were classified into six categories: normal tissue (950 ROIs), hyperplastic polyps (545 ROIs), tubular adenoma with high-grade dysplasia (454 ROIs), tubular adenoma with low-grade dysplasia (3,618 ROIs), tubulo-villous adenoma with high-grade dysplasia (916 ROIs), and tubulo-villous adenoma with low-grade dysplasia (2,186 ROIs). The dataset was split into training and test sets (6,068:2,601 ROIs). To evaluate the resolution sensitivity and adaptability of pathology foundation models, we conducted linear head fine-tuning, and SimpleShot learning using five different input resolutions: 224, 448, 672, 896, and 1,120 pixels. Additionally, we visualized the multi-head attention distributions of the pretrained models and the fine-tuned Transformer layers across different resolutions to explore attention dynamics at varying scales.

**CAMEL colorectal adenoma screening (2 classes, CAMEL).** The screening dataset comprises 15,403 ROI images extracted from 177 colorectal slides from the Department of Pathology, Chinese PLA General Hospital. The original resolution of the images is 1,280 × 1,280 pixels and we resized it to 224 × 224 pixels during the experiments. The cohort consisted of 8,450 adenoma-containing ROIs and 6,953 normal tissue ROIs.To ensure rigorous evaluation, the dataset is partitioned into training (10,782 ROIs) and test (4,621 ROIs) subsets.

**Colorectal epithelial tumors and proliferative lesion classification (3 classes and 5 classes, IMP-CRS2024 and BOW-5cls).** This dataset includes both the IMP-CRS2024 public dataset and a custom-built dataset. We random selected 1132 colorectal WSIs from the IMP-CRS2024 training dataset and used the officail test dataset (900 WSIs) for evaluation, which were labeled into three categories: non-tumorous lesions (484 WSIs), low-grade lesions (1004 WSIs), and high-grade lesions (544 WSIs). For the slide-level classification task, 10× magnification images were used as model input.

Self-built dataset includes 1,686 colorectal WSIs collected from LZ, annotated into six categories: class 0 (normal, 522 WSIs), class 1 (hyperplastic polyps, 130 WSIs), class 2 (LIN, 379 WSIs), class 3 (HIN, 163 WSIs), class 4 (adenocarcinoma, 492 WSIs). The dataset was divided into training, validation and test sets (1,180:169:337 WSIs). For the slide-level classification task, 10× magnification images were used as model input.

**Early esophageal cancer screening (2 classes, ESO-2cls).** This task utilized a self-built dataset consisting of 860 WSIs collected from ZN, LZ, and SZT, with all cases histologically classified as stage T1a or T1b according to the WHO criteria. The WSIs were grouped into two categories: Class 0 included squamous epithelial papilloma and chronic esophagitis (415 WSIs); Class 1 included LIN, HIN, SCC, and esophageal adenocarcinoma (445 WSIs); A five-fold cross-validation strategy was adopted, with each fold further split into training, validation, and test sets (550:138:172 WSIs). All classification tasks were performed using 20× magnification during the MIL preprocessing stage.

**Esophageal epithelial neoplasia classification (4 classes, ESO-4cls).** This task was conducted on an self-built dataset comprising 860 WSIs collected from three medical centers (ZN, LZ, and SZT). The WSIs were categorized into five classes: Class 0 included non-neoplastic cases such as squamous epithelial papilloma and chronic esophagitis (415 WSIs); Class 1 included LIN (29 WSIs); Class 2 included HIN (150 WSIs); and Class 3 consisted of carcinoma (266 WSIs). A five-fold cross-validation scheme was adopted, with each fold further split into training, validation, and test sets (550:138:172 WSIs). All models were trained and evaluated using input patches at 20× magnification during the MIL preprocessing stage.

**Differentiation between keratinizing and non-keratinizing subtypes of esophageal SCC (2 classes, ENSOKRT)**. A total of 338 WSIs were collected from three medical centers (ZN, LZ, and SZT) as part of a self-built dataset. The dataset was divided into two classes: Class 0 consisted of keratinizing ESCC (167 WSIs), and Class 1 comprised non-keratinizing ESCC (171 WSIs). A five-fold cross-validation protocol was employed, with each fold further split into training, validation, and test sets (216:55:67 WSIs). All input patches were processed at 20× magnification during MIL-based classification.

**Differentiation between esophageal SCC and adenocarcinoma (2 classes, ESO-AS).** This dataset comprised 349 WSIs collected from three medical centers (ZN, LZ, and SZT), annotated into two categories: Class 0 consisted of esophageal SCC (272 WSIs) and Class 1 consisted of esophageal adenocarcinoma (77 WSIs). A five-fold cross-validation strategy was used, with each fold split into training, validation, and test sets (223:56:70 WSIs). All slides were processed at 20× magnification during MIL-based classification.

**Intestinal metaplasia gland segmentation (IM-S).** This dataset comprised 85 biopsy-derived WSIs of intestinal metaplasia collected from FJ, which were cropped into 1,135 image patches with size of 512 × 512 pixels. Each patch was meticulously annotated at the pixel level by pathologists, distinguishing intestinal metaplastic glands from non-intestinal metaplastic glands. The dataset was divided into training, validation, and test sets (794:114:227 ROIs), following five replicate experiments. For the gland segmentation task, 10× magnification images were used as model input.

**Gastric tumor region segmentation (STESD-S).** This dataset comprised 60 endoscopic submucosal dissection (ESD) gastric tumor slides collected from ZN and FJ. The slides were divided into 4,455 image patches of size 512 × 512 pixels. Each patch was meticulously annotated at the pixel level by pathologists to delineate tumor and non-tumor regions. The dataset was split into training, validation, and test sets (3,120:445:890 ROIs), following five replicate experiments. For the tumor region segmentation task, 10× magnification images were used as model input.

**CRAG colorectal gland segmentation (CRAG)**. This public dataset contains 213 images taken from 38 H&E stained WSIs of colorectal adenocarcinoma. we used the officially provided training–test split (173:40 ROIs), following five replicate experiments. All images mostly have a size of 1512 × 1516 with pixel-level gland annotations.

**Gastric cancer staging prediction (4 classes, STM-TNM).** This dataset comprised 300 WSIs collected from ZN. Based on the 8th edition of the AJCC TNM staging system, the WSIs were categorized into four stages: class 0 (stage I, 100 WSIs), class 1 (stage II, 74 WSIs), class 2 (stage III, 106 WSIs), and class 3 (stage IV, 20 WSIs). The dataset was divided into training, validation, and test sets (192:48:60 WSIs), following a five-fold cross-validation protocol. For the slide-level classification task, 20× magnification images were used as model input.

**Colorectal cancer staging prediction (4 classes, BOW-TNM).** This dataset comprised 995 WSIs collected from ZN. Based on the 8th edition of the AJCC TNM staging system, the WSIs were categorized into four stages: class 0 (stage I, 194 WSIs), class 1 (stage II, 343 WSIs), class 2 (stage III, 340 WSIs), and class 3 (stage IV, 194 WSIs). The dataset was divided into training, validation, and test sets (637:159:199 WSIs), following a five-fold cross-validation protocol. For the slide-level classification task, 20× magnification images were used as model input.

**Differentiation between poorly differentiated adenocarcinoma and poorly differentiated SCC (2 classes, LA-LS).** Differentiating between poorly differentiated adenocarcinoma and poorly differentiated SCC in the gastrointestinal tract presents a major diagnostic challenge. This dataset comprised 384 WSIs collected from four medical centers (ZN, LZ, SZT, FJ), annotated into two categories: Class 0 (poorly differentiated adenocarcinoma, 236 WSIs) and Class 1 (poorly differentiated SCC, 148 WSIs). A five-fold cross-validation strategy was used, with each fold split into training, validation, and test sets (246:61:77 WSIs). All slides were processed at 20× magnification during MIL-based classification.

**Precancerous lesions and reactive hyperplasia (3 classes, LHN-3cls).** The dataset comprised 462 WSIs collected from four medical centers (ZN, LZ, SZT, FJ), annotated into two classes: Class 0 consisted of reactive hyperplasia (160 WSIs), Class 1 consisted of LIN (93 WSIs), and Class 2 consisted of HIN (209 WSIs). A five-fold cross-validation scheme was employed, with each fold further split into training, validation, and test sets (296:74:92 WSIs). All slides were processed at 20× magnification during MIL-based classification.

**Differentiation between signet-ring cell carcinoma and histiocytes (2 classes, R-X).** This dataset consisted of 384 WSIs collected from four medical centers (ZN, LZ, SZT, FJ), annotated into two categories: Class 0 (histiocytes, 182 WSIs) and Class 1 (signet-ring cell carcinoma, 202 WSIs). A five-fold cross-validation protocol was applied, with each fold further divided into training, validation, and test sets (246:61:77 WSIs). All slides were processed at 20× magnification during MIL-based classification

**Molecular status prediction (2 classes, PD-L1, P53, HER2).** To assess the capability of the model in predicting molecular markers from routine histopathology, we constructed three in-house datasets from ZN, targeting PD-L1, P53, and HER2 expression status. The PD-L1 dataset consisted of 751 WSIs (positive:negative = 483:268 WSIs) from the stomach. Here, positive cases were defined as CPS (Combined Positive Score) > 0, indicating detectable PD-L1 expression, while negative cases (CPS = 0) showed no PD-L1 expression. The P53 dataset included 710 WSIs (mutant-type:wild-type = 361:349 WSIs) from the esophagus, and the HER2 dataset comprised 399 WSIs (positive:negative = 92:307 WSIs) from the stomach. For each task, five-fold cross-validation was performed, with each fold split into training, validation, and test sets: PD-L1 (479:121:151 WSIs), P53 (454:114:142

WSIs), and HER2 (255:64:80 WSIs). All images were processed at 20× magnification during MIL-based classification.

**Microsatellite instability prediction (2 classes, MSI).** An self-built dataset comprising 970 surgical WSIs from ZN was used to evaluate MSI prediction performance in gastrointestinal cancers, including both gastric and colorectal specimens. This dataset was categorized into two groups: Class 0 (microsatellite instable) and Class 1 (microsatellite stable). A five-fold cross-validation strategy was employed, with each fold divided into training, validation (582:194:194 WSIs). All slides were processed at 20× magnification for MIL-based classification.

**Gene mutation prediction (2 classes, TCGA-COAD-MUC16, TCGA-COAD-TTN, TCGA-READ-TTN, TCGA-READ-NRAS).** This study performed gene mutation prediction based on gastrointestinal-related datasets from The Cancer Genome Atlas (TCGA), including COAD (colon adenocarcinoma, 403 WSIs), and READ (rectum adenocarcinoma, 128 WSIs). We focused on the three frequent driver genes (*MUC16* in TCGA-COAD , *TTN* in TCGA-COAD, *NRAS* in TCGA-READ, and *TTN* in TCGA-READ) in these cancer types. A five-fold cross-validation strategy was used, with each fold split into training, validation, and test sets (258:64:81 WSIs for TCGA-COAD, and 82:20:26 WSIs for TCGA-READ). All slides were processed at 20× magnification during MIL-based classification.

**Survival prediction(TCGA-COAD-SUR, TCGA-READ-SUR).** For survival outcome modeling, we curated a dataset from 408 from colon adenocarcinoma (COAD), and 153 from rectum adenocarcinoma (READ). A five-fold cross-validation strategy was used, with each fold split into training, validation, and test sets (261:65:82 WSIs for COAD and 98:24:31 WSIs for READ). All slides were processed at 20× magnification for MIL-based classification.

**Prospective multi-center study for early cancer screening**

According to the 5th edition of the WHO of the Digestive System, we defined positive samples as those diagnosed with LIN, HIN, or confirmed malignant tumors. All other samples, including non-neoplastic lesions and benign polyps, were labeled as negative. A prospective validation study was conducted across nine hospitals, including ZN, LZ, FJ, SZT, Tsinghua Changgung Hospital (CG), Chongqing University Affiliated Three Gorges Hospital (SX), The First Affiliated Hospital of Nanchang University (NC), Ningbo Clinical Pathology Diagnosis Center (NB), and Longgang Central Hospital of Shenzhen (LG), which represent a wide geographic distribution across eastern, southern, western, and northern China. Each hospital tested approximately 1,000 biopsy slides, yielding a total of 11,224 WSIs, among which 657 were positive. Slides distribution across the hospitals is summarized in Fig. 5 and Supplementary Table 47.

**Computing hardware and software**

All experiments and analyses were implemented in Python 3.8.13 with PyTorch 2.0.0 (CUDA 11.7). The computational framework is fully reproducible using open-source libraries and codebases as follows. For Digepath pretraining, we modified the Vision Transformer from the timm 0.9.2 library (https://huggingface.com) as the encoder backbone and integrated it with the original DINOv2 self-supervised algorithm, with pretraining executed on 8 × 80GB NVIDIA A800 GPUs. Downstream tasks were performed on a single 24GB NVIDIA RTX 4090 GPU. WSI processing relied on OpenSlide 4.3.1, openslide-python 1.2.0, opensdpc (GitHub: WonderLandxD/opensdpc), and the CLAM framework (GitHub: mahmoodlab/CLAM). Benchmark visual encoders included CTransPath (https:// github.com/Xiyue-Wang/TransPath), UNI (https://github.com/mahmoodlab/UNI), Gigapath (https://github.com/prov-gigapath/prov-gigapath), and Conch-V1.5 (https://github.com/mahmoodlab/TITAN). Weakly supervised multiple instance learning (MIL) models were adapted from the MIL_BASELINE codebase (https://github.com/lingxitong/MIL_BASELINE), while semantic segmentation utilized the TransUNet implementation (https://github.com/Beckschen/TransUNet).

Evaluation protocols for linear probing and prototypical networks were based on the UNI codebase (https://github.com/mahmoodlab/UNI). Visualization workflows employed Pillow 9.3.0, Matplotlib 3.7.1, and Seaborn 0.12.2. All referenced code repositories are publicly accessible through their respective GitHub URLs provided in the manuscript.

**Data availability**

TCGA data consisting of WSIs and labels can be accessed through the NIH genomic data commons (https://portal.gdc.cancer.gov).

CRC-100K data can be accessed through the Zenodo database (https://zenodo.org/record/1214456).

CAMEL data can be accessed through the github link (https://github.com/ThoroughImages/CAMEL).

CRAG data can be accessed through the github link (https://github.com/XiaoyuZHK/CRAG-Dataset_Aug_ToCOCO).

UNITOPATHO data can be accessed through the ieee-dataport database (https://ieee-dataport.org/open-access/unitopatho).

IMP-CRS data can be accessed through the link (https://rdm.inesctec.pt/dataset/nis-2023-008).

The private pathological images can be obtained by contacting the corresponding author (heyh@sz.tsinghua.edu.cn) for scientific research purposes.

**Code availability**

Code and model weights for Digepath can be accessed later for academic research purposes at https://github.com/lingxitong/Digepath. We have documented all technical deep learning methods and software libraries used in the study while ensuring that the paper is accessible to the broader clinical and scientific audience.

**Ethics**

This retrospective study received ethical approval for clinical/scientific research projects under medical ethics committee, Zhongnan Hospital of Wuhan University (No. [2025010K]), Fujian Provincial Hospital (No. K2024-09-047), Liuzhou People's Hospital (No. KY-2025-232), and the Second Affiliated Hospital of Southern University of Science and Technology (No. 2025-045-02). Prior to undergoing biopsies and pathological examinations, all patients involved in the study provided informed consent by signing the necessary documents.

**Acknowledgements**

We thank Shenzhen Shengqiang Technology Co., Ltd. for providing slide scanners; H3C Technologies Co., Ltd. for providing the training servers; This work was supported in part by National Natural Science Foundation of China (82430062), the Shenzhen EngineeringResearch Centre (XMHT20230115004), the Jilin FuyuanGuan Food Group Co., Ltd., Fujian Provincial Science and Technology Innovation Joint Funds (grant no. 2024Y96010076), and the Fujian Provincial Natural Science Foundation of China (grant no. 2024J011006).

**Author contributions**

L.H.Z., X.T.L., M.X.O.Y., Y.H.H., and S.F.T. conceived the study and designed the experiments. L.H.Z., X.T.L., M.X.O.Y., X.P.L., S.F.T., L.X.X., H.Q.L. collected the data for self-supervised learning. L.H.Z., X.T.L., M.X.O.Y. Y.H.H. performed model development for self-supervised learning. L.H.Z., X.T.L., M.X.O.Y., M.X.F, F.L.F., M.M.Z., M.X.Z., Y.B.J., Q.M.H., Y.Z.W., and J.R.C. built code bases for all downstream tasks. L.H.Z., X.T.L., M.X.O.Y., X.P.L., Z.Q.C., L.M.L., S.D., Q.H., Y.X., J.M.L., S.M.L., S.X., L.X.X., Z.H.P., and H.Q.L. collected the datasets all downstream tasks. T.G., Q.H. J.R.C. and H.Q.L. performed quality control of the codebase and the results. L.H.Z., X.T.L., M.X.O.Y., Y.H.H., and S.F.T. performed the experiments analysis. L.H.Z., X.T.L., X.P.L., M.X.O.Y., M.X.F., F.L.F., M.M.Z., M.X.Z., Y.B.J., Y.H.H., and S.F.T. interpreted the results and provided feedback on the study. L.H.Z., X.T.L., M.X.O.Y., and X.P.L. prepared the manuscript. Y.H.H., and S.F.T. supervised the research. L.H.Z., X.T.L., M.X.O.Y., X.P.L., M.X.F. and F.L.F. have accessed and verified data. All authors have read and agreed to publish the paper.

**Competing interests**

All authors declare no competing interests.

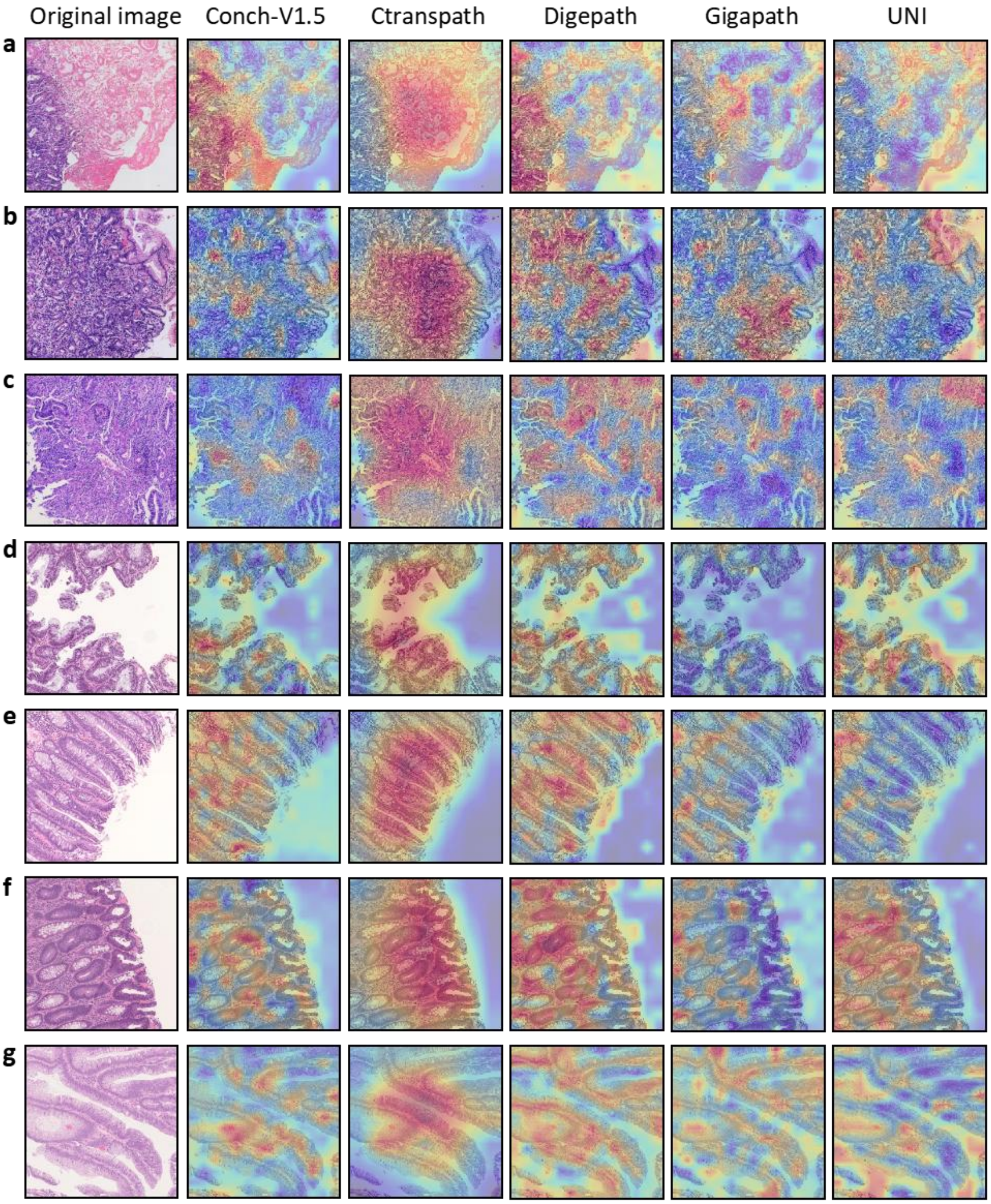

**Extended Data Figure 1 | ROI visualizations across models. a–c.** Visualizations of five models on STLC. **d–g.** Visualizations of five models on UNITOPATHO.

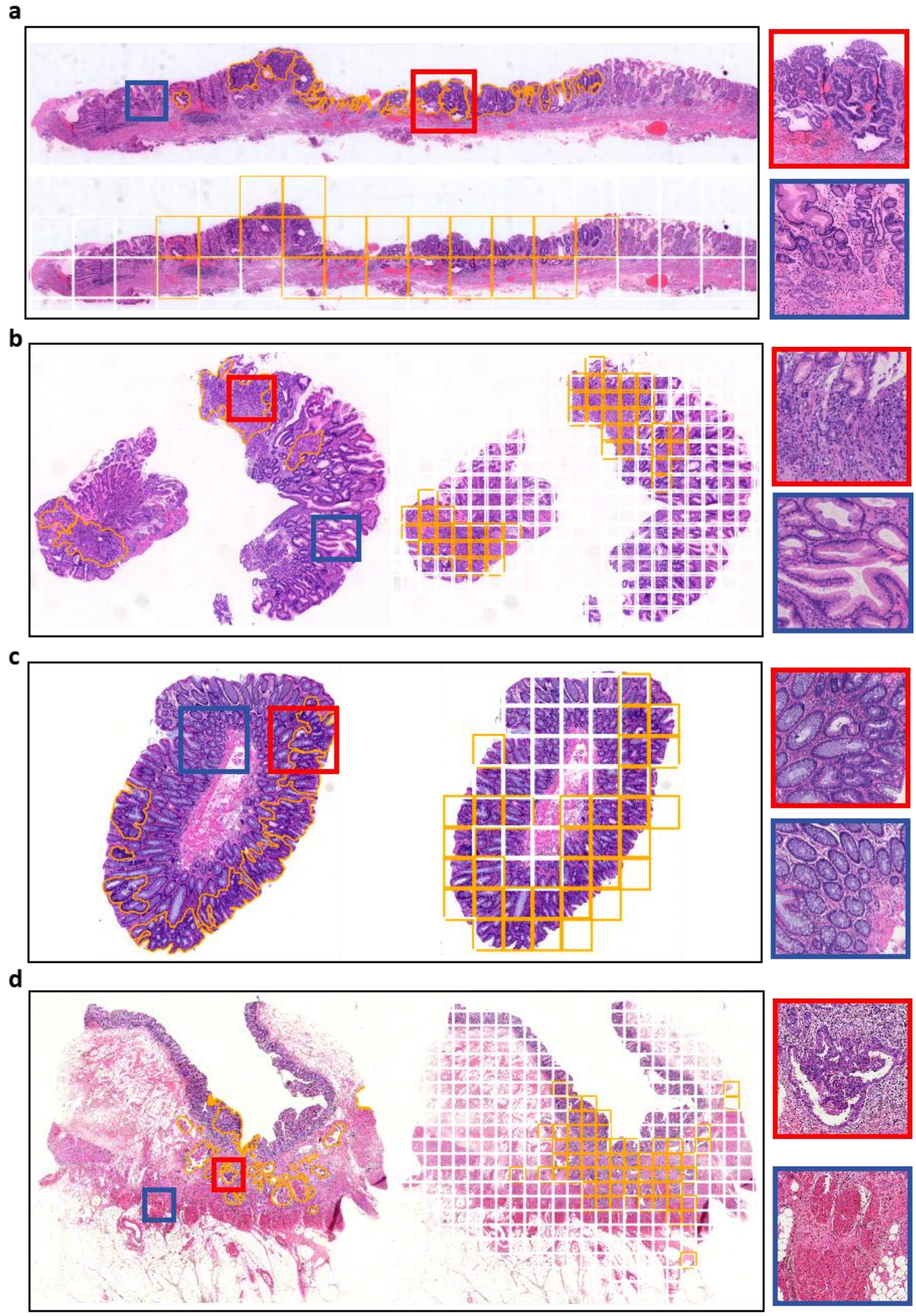

**Extended Data Figure 2 | ROI-based tumor region detection visualizations.** For panels a–d, the top-left inset shows pathologist-annotated tumor regions; the bottom-left shows model predictions (orange boxes: tumor; white boxes: non-tumor); the top-right and bottom-right insets are zoomed views of the red and blue boxes, respectively. **a.** ESD specimen (gastric). **b.** Biopsy specimen (gastric). **c.** Surgical specimen (gastric). **d.** Surgical specimen (intestinal).

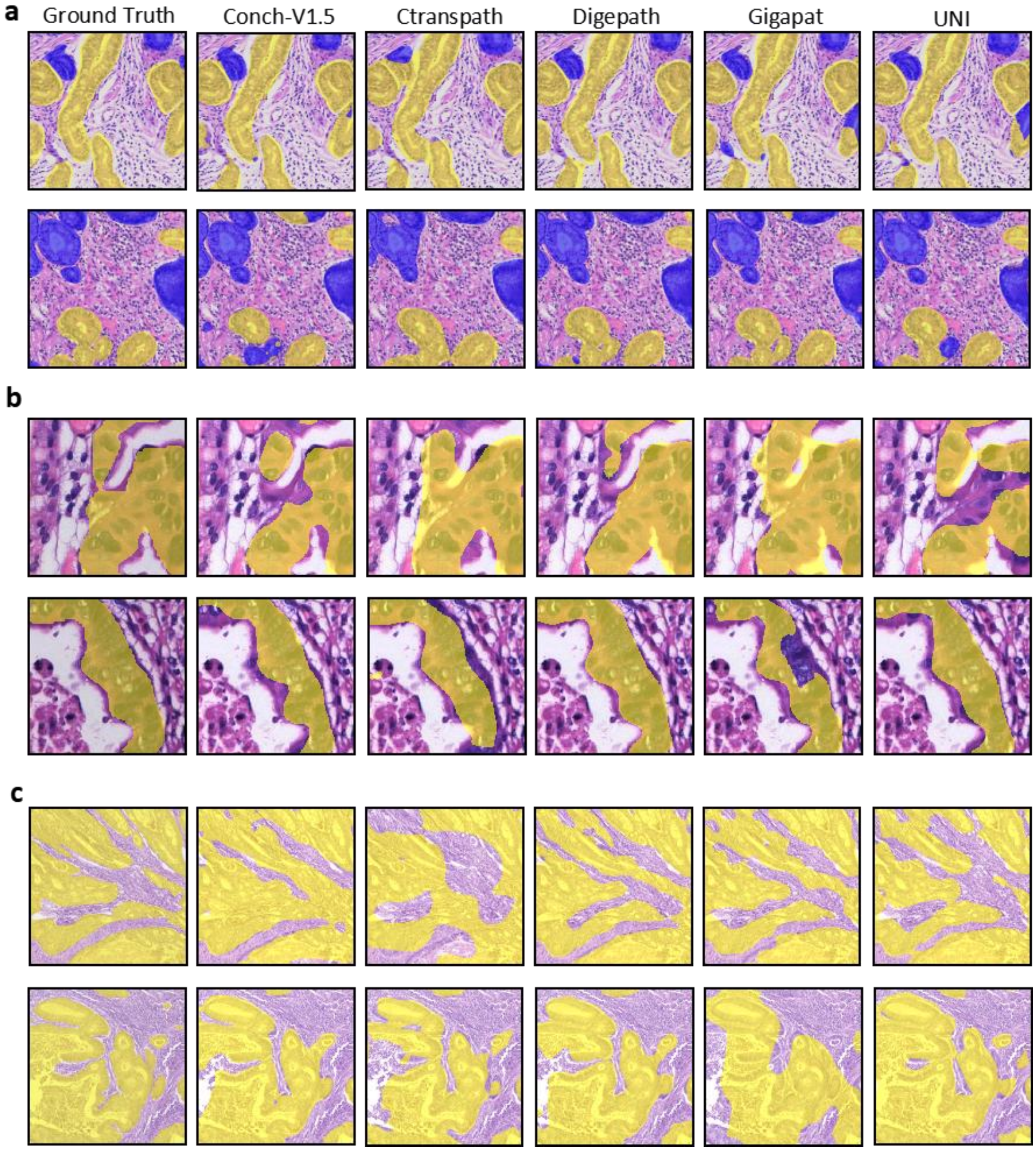

**Extended Data Figure 3 | Visualizations of segmentation task. a.** Visualization of Digepath on intestinalized/non- intestinalized gland segmentation. **b.** Visualization of Digepath on ESD tumor region segmentation. **c.** Visualization of Digepath on gland and tumor segmentation using the public data (CRAG).

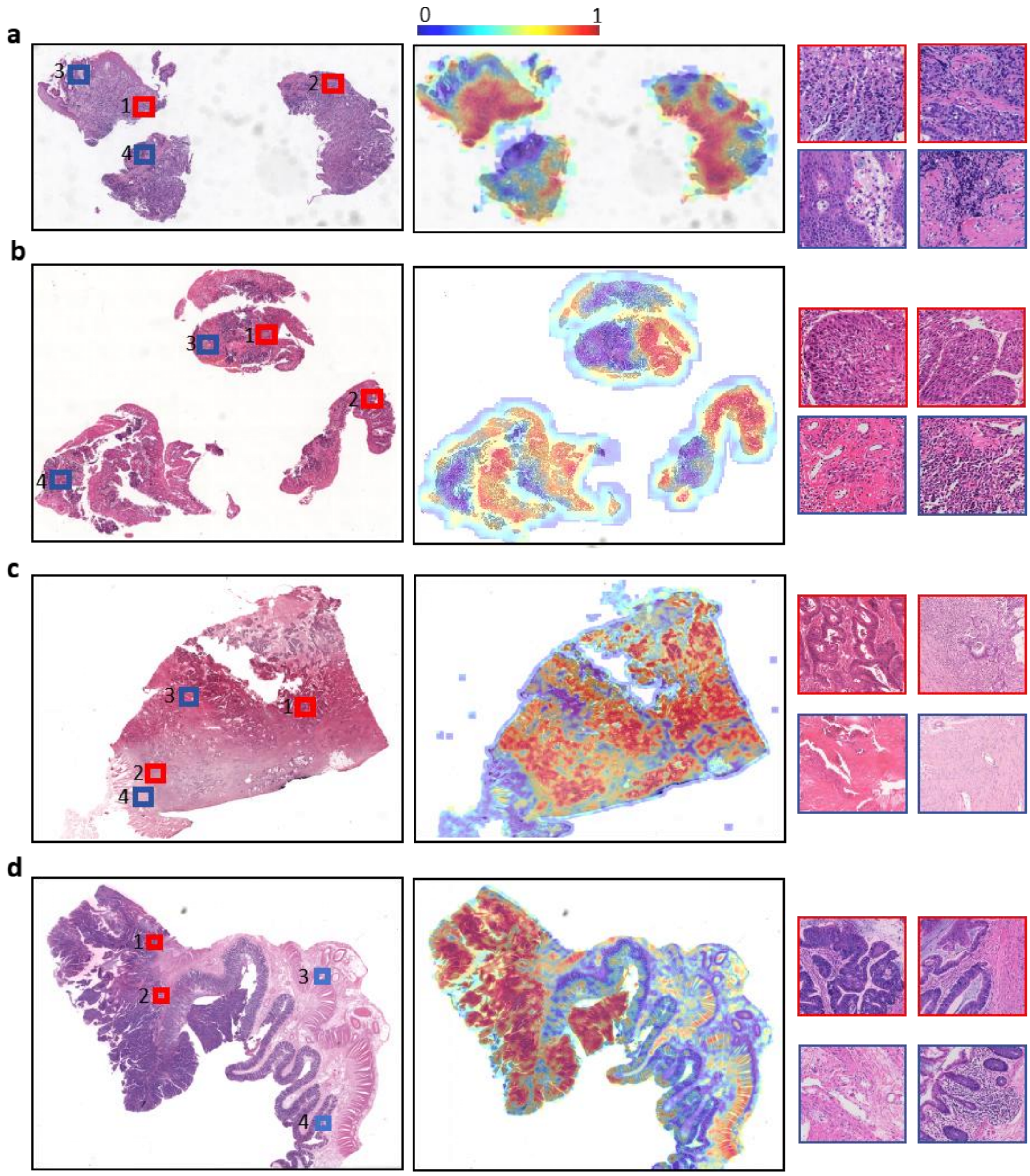

**Extended Data Figure 4 | Attention-based visualizations on challenging cases.** For panels a–d, the left image shows the original slide (red box: pathologist-annotated tumor; blue box: pathologist-annotated non-tumor), the center image shows model-predicted tumor regions, and the right insets show zoomed views of the red and blue boxes. **a.** Poorly differentiated adenocarcinoma. **b.** Poorly differentiated SCC. **c.** Stomach TNM staging task. **d.** Intestinal TNM staging task.

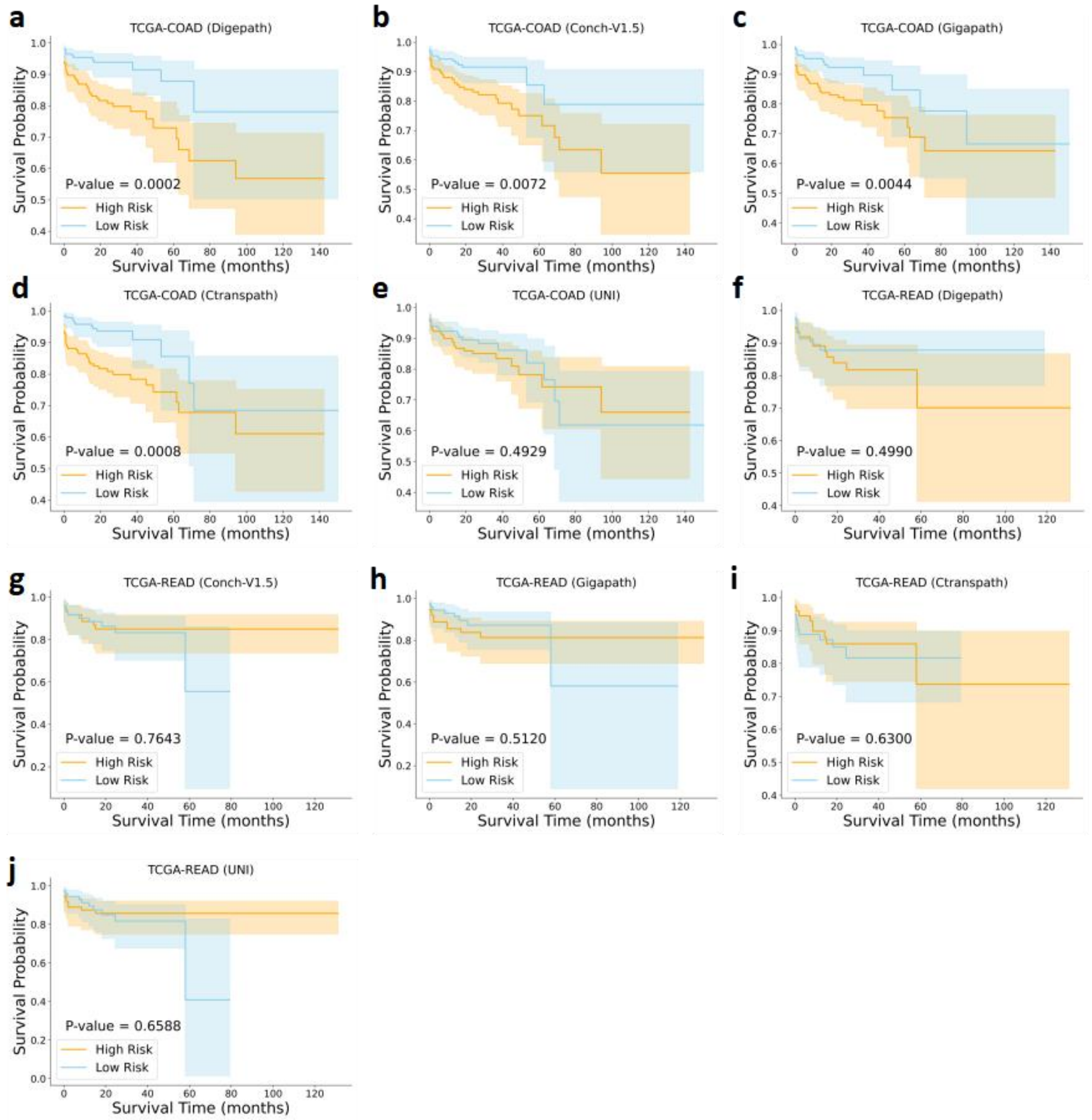

**Extended Data Figure 5 | Kaplan–Meier curves for survival prediction. a–e**. K-M curves across models on TCGA- COAD (82 WSIs). **f–j**. K-M curves across models on TCGA- READ (31 WSIs).

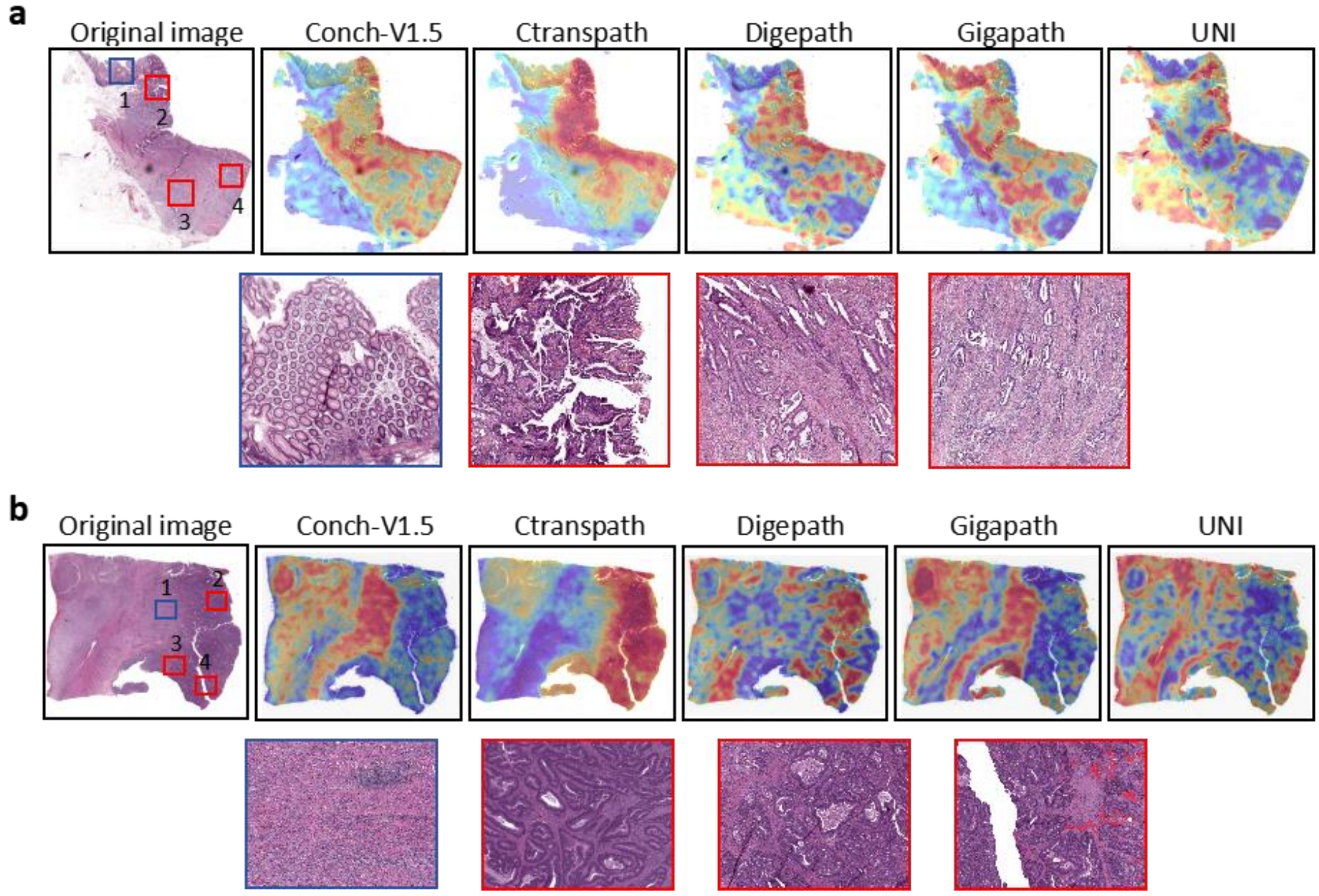

**Extended Data Figure 6 | Survival visualizations across various models. a–b.** the first image in the top row is the original slide (red box: pathologist-annotated tumor; blue box: pathologist-annotated non-tumor), followed by heatmap visualizations from each model; the bottom row shows zoomed views of regions 1–4 from the first image.

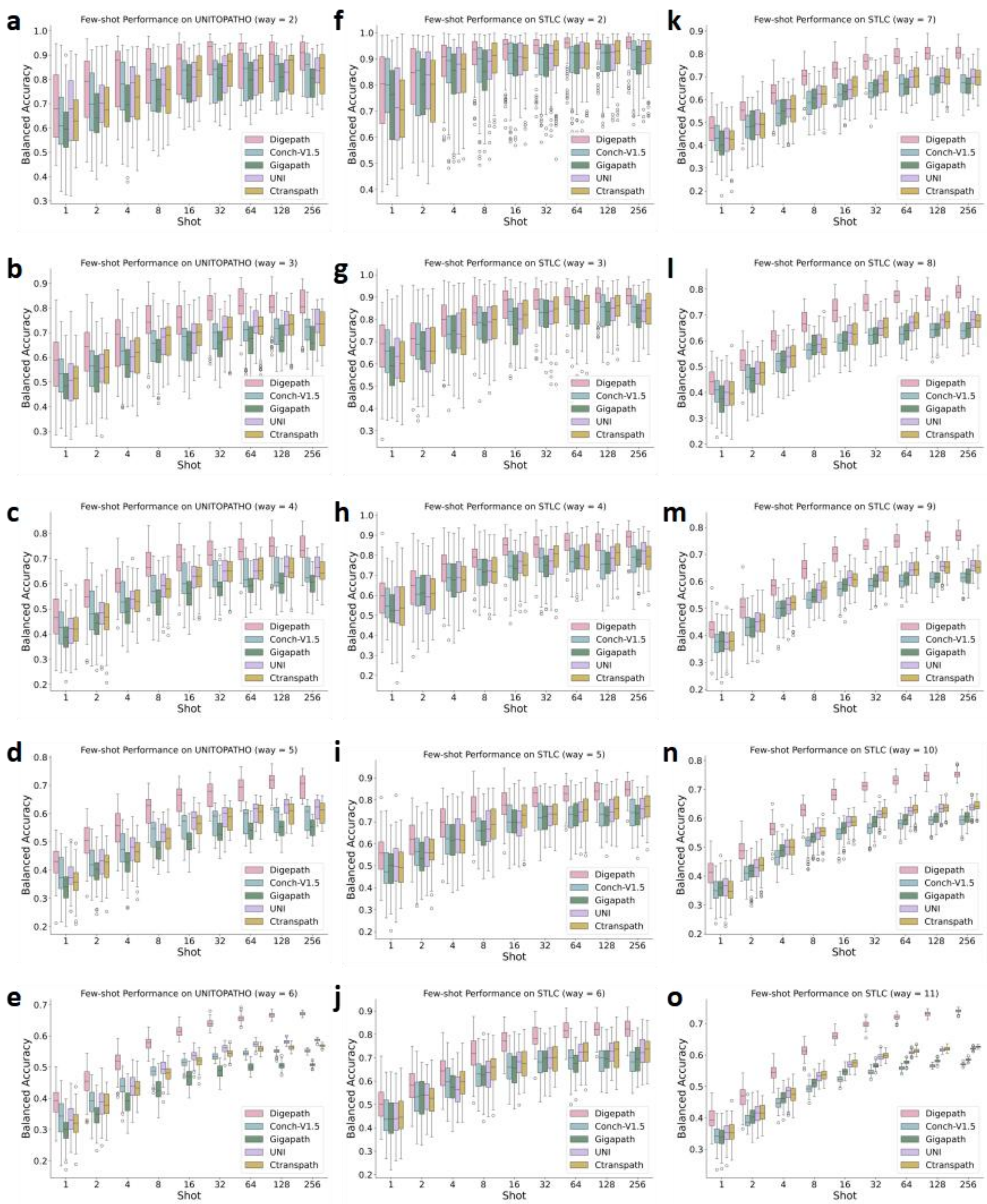

**Extended Data Figure 7 | Full-way-few-shot performance comparison. a–e**. Few-shot learning performance across models as the way number increases from 2 to 6 on UNITOPATHO. **f–o.** Few-shot learning performance across models as the way number increase from 2 to 11 on STLC.

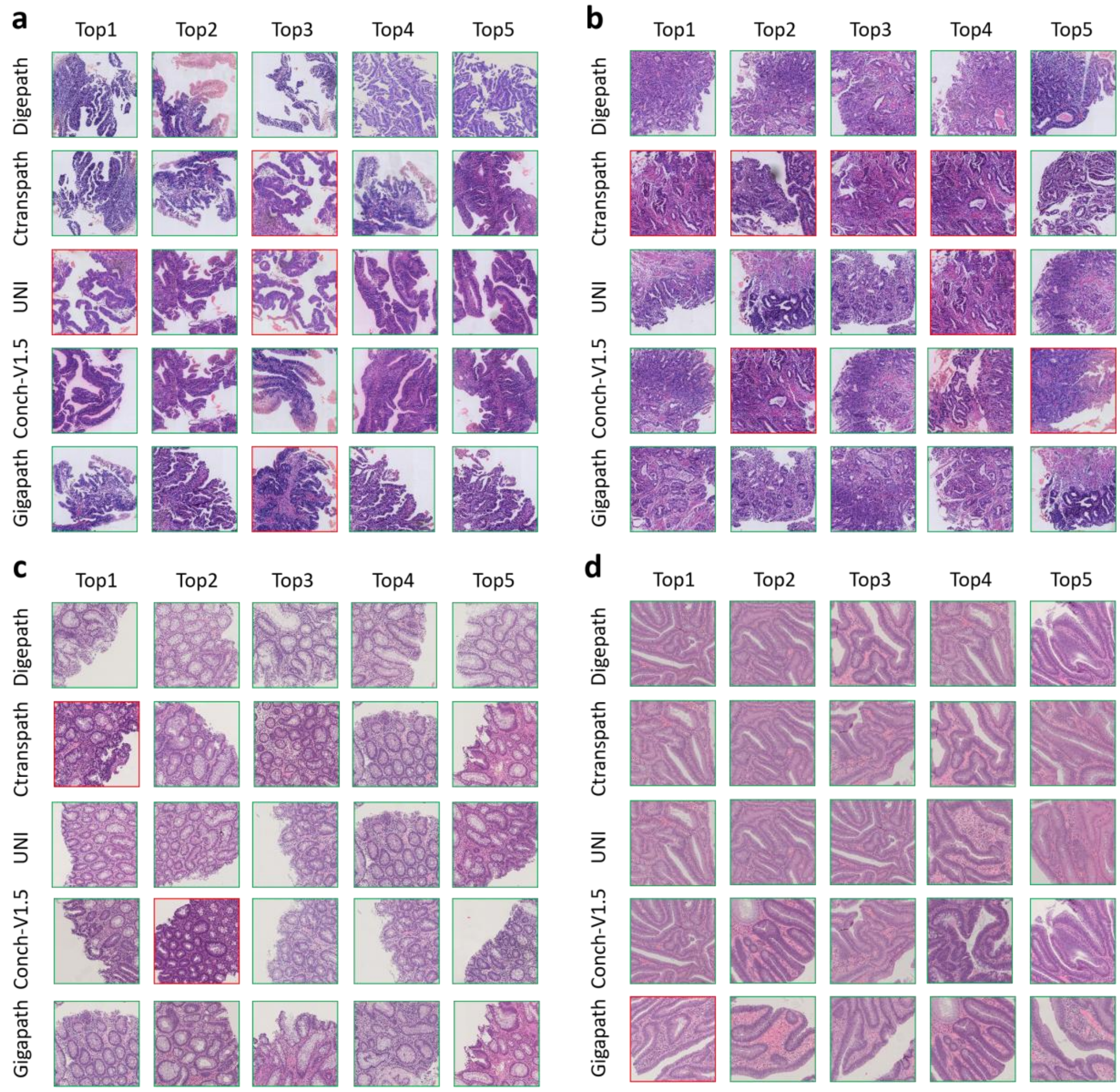

**Extended Data Figure 8 | Image retrieval performance comparison across models.** **a.** A comparative visualization of the 5 highest similarity images to the papillary class prototype, as retrieved by distinct models on STLC. **b.** A comparative visualization of the 5 highest similarity images to the HGIN class prototype, as retrieved by distinct models on STLC. **c.** A comparative visualization of the 5 highest similarity images to the low-grade tubular adenoma class prototype, as retrieved by distinct models on UNITOPATHO. **d.** A comparative visualization of the 5 highest similarity images to the high-grade tubular adenoma class prototype, as retrieved by distinct models on UNITOPATHO.

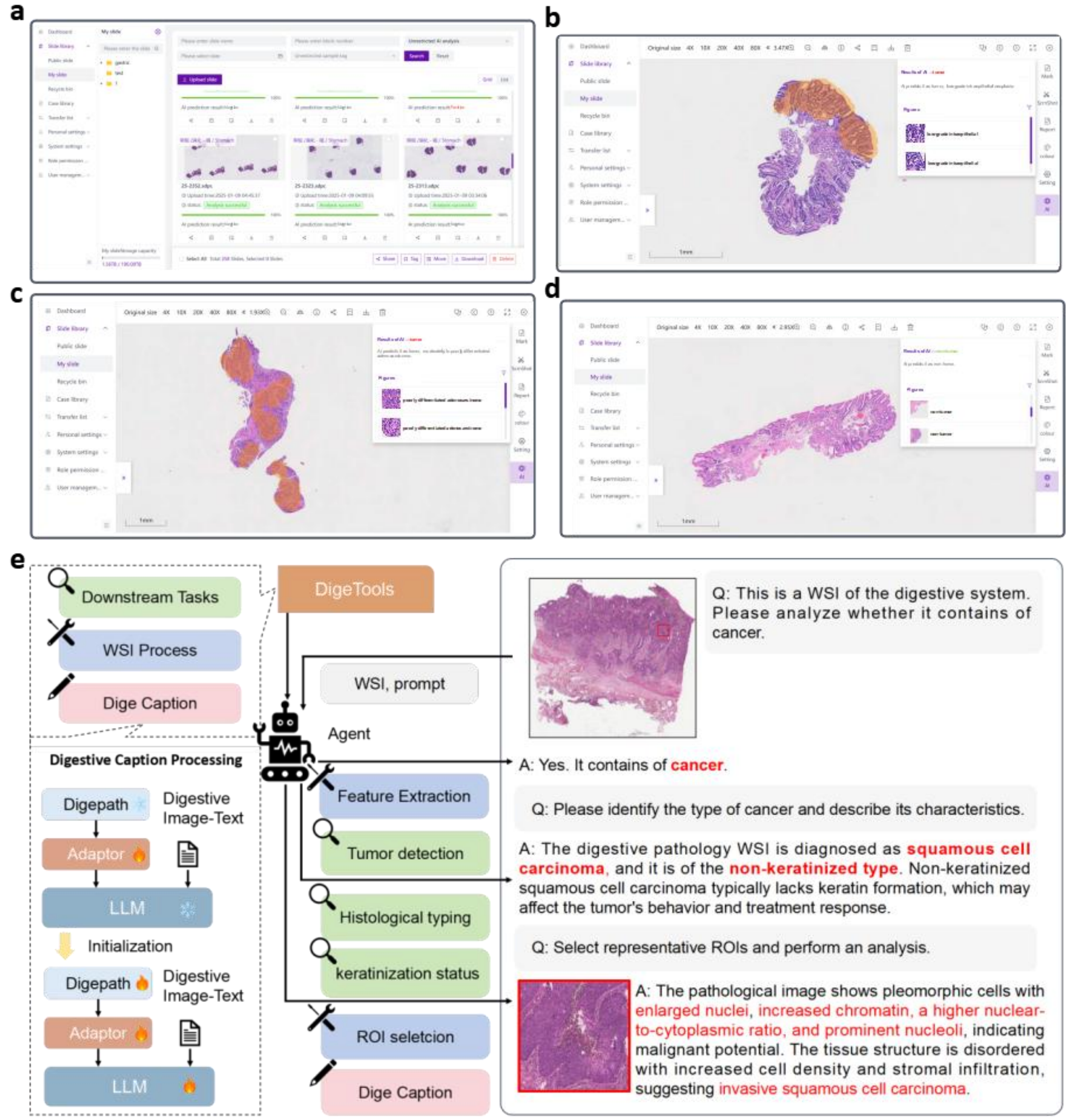

**Extended Data Figure 9 | Early-cancer screening module interface and test results. a.** Screenshot of the early-cancer screening software interface. **b–d.** Test examples, **b.** Low-grade epithelial neoplasia. **c**. Poorly differentiated adenocarcinoma. **d.** Non-tumor case. **e.** The end-to-end pipeline of the agent system for GI pathology proposed. Taking a WSI of digestive tissue together with a user text prompt as inputs, the system leverages the DigeTools library to sequentially perform pathology feature extraction, cancer detection, subtype identification, and ROI-level report delivery through multi-turn dialogue.